\documentclass{sig-alternate-10pt}
\usepackage{times}
\usepackage{cite}	
\usepackage{url}
\usepackage{multirow}	
\usepackage{xspace}

\long\def\com#1{}

\long\def\xxx#1{}

\long\def\abbr#1#2{#1}			

\newcommand{\abcite}[2]{\abbr{\cite{#1}}{\cite{#1,#2}}}

\newcommand{\blind}[1]{#1}		

\renewenvironment{itemize}{
   \begin{list}{\labelitemi}{
     \setlength{\topsep}{0.5ex}
     \setlength{\itemsep}{-0pt}
     \setlength{\itemindent}{0pt}
     \setlength{\leftmargin}{\labelwidth}
     \addtolength{\leftmargin}{-8pt}}
}{\end{list}}

\def\TNG{T{\bf ng}\xspace}	
\def\Tng{T{\em ng}\xspace}
\def\tng{T{\em ng}\xspace}

\begin{document}

\title{ Flow Splitting with Fate Sharing in
	\\ a Next Generation Transport Services Architecture
       	\\ {\tt UNPUBLISHED DRAFT}}

\blind{
\numberofauthors{2}
\author{
        \alignauthor
        Janardhan Iyengar \\
        \affaddr{Franklin and Marshall College} \\
        \email{jiyengar@fandm.edu} \and
        \alignauthor
	Bryan Ford	\\
	\affaddr{Max Planck Institute for Software Systems}\\
	\email{baford@mpi-sws.org}
}
}

\maketitle

\begin{abstract}

The challenges of optimizing end-to-end performance
over diverse Internet paths
has driven widespread adoption of in-path optimizers,
which can destructively interfere with
TCP's end-to-end semantics and with each other,
and are incompatible with end-to-end IPsec.
We identify the architectural cause of these conflicts
and resolve them in \tng,
an experimental next-generation transport services architecture,
by factoring congestion control
from end-to-end semantic functions.
Through a technique we call {\em queue sharing},
\tng enables in-path devices to interpose on, split, and optimize
congestion controlled flows
without affecting or seeing
the end-to-end content riding these flows.
Simulations show that \tng's decoupling
cleanly addresses several common performance problems,
such as communication over lossy wireless links
and reduction of buffering-induced latency on residential links.
A working prototype and several incremental deployment paths
suggest \tng's practicality.

\end{abstract}

\section{Introduction}

Ever since TCP congestion control was introduced~\cite{jacobson88congestion},
we have found reasons to tweak it within the network.
Performance enhancing proxies (PEPs)~\cite{rfc3135}
improve TCP's poor performance
over loss-prone wireless links~\cite{yavatkar94improving},
intermittent mobile links~\cite{bakre97implementation},
and high-latency satellite links~\cite{cisco04rate}.
Due to their effectiveness and ease of deployment,
PEPs now form the technical foundation
of a booming \$1 billion WAN optimization market~\cite{mcgillicuddy09wan},
and are joining the growing class of middleboxes
such as firewalls~\cite{rfc2979}, NATs~\cite{rfc3022},
and flow-aware routers~\cite{roberts03next}
pervading the Internet.

PEPs are in theory compatible with the end-to-end principle~\cite{
	saltzer84endtoend},
which argues that reliability mechanisms need to be end-to-end
but explicitly allows for in-network mechanisms to enhance performance
as long as they do not replace end-to-end reliability checks.
Because the Internet's {\em architecture}
lumps congestion control with end-to-end reliability
in the transport layer,
however,
PEPs in the path cannot affect one function without interfering with the other.
Many PEPs violate fate-sharing~\cite{clark88design}
by introducing ``hard state'' in the network,
causing application-visible failures if a PEP crashes.
All PEPs are incompatible with
transport-neutral security mechanisms such as end-to-end IPsec~\cite{rfc4301},
which prevent the PEP from seeing the relevant transport headers.

\com{	too peripheral for intro  -baf
Besides PEPs, ubiquitous network middleboxes
such as firewalls~\cite{rfc2979}, NATs~\cite{rfc3022},
and flow-aware routers~\cite{roberts03next}
routinely interact with transport headers to obtain the information they need
to implement security, connectivity, or traffic management policies,
resulting in the current problem that
new transports~\abcite{rfc4340,rfc4960}{kohler06dccp}
cannot traverse most middleboxes~\cite{rosenberg08udp}.
Middleboxes were not conceived as part of the original Internet architecture, 
and---as their deployment continues unabated
for policy and/or performance reasons---
they still haven't been accepted into it,
leading to the kinds of problems faced by PEPs described above.
}

\begin{figure}[t]
\centering
\includegraphics[width=0.48\textwidth]{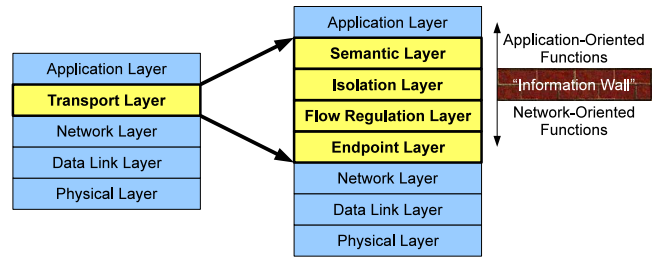}
\caption{\tng Architecture Layering}
\label{fig-layers}
\end{figure}

Our novel solution to this architectural dilemma
is to refactor the transport layer
so that PEPs {\em can} cleanly interpose on and optimize
congestion control behavior,
without interfering with,
or even seeing the protocol headers for,
end-to-end functions such as reliability.
We develop this approach in the context of \tng,
an experimental next-generation transport
that builds on ideas introduced earlier~\cite{
	ford07structured,ford08breaking}
to address a broader class of transport issues.

\tng breaks transports into four layers,
shown in Figure~\ref{fig-layers}.
\tng's {\em Semantic Layer} implements end-to-end abstractions
such as reliable byte streams;
its optional {\em Isolation Layer} protects upper end-to-end layers
from in-path interference;
its {\em Flow Regulation Layer} factors out performance concerns
such as congestion control to enable performance management by PEPs;
and its {\em Endpoint Layer} factors out endpoint naming concerns
such as port numbers
to enable clean NAT/\nolinebreak[0]firewall traversal~\cite{
	ford05p2p}.
We make no claim
that \tng represents ``the ideal architecture,''
but use it here only to develop a cleaner solution
to the problem of PEPs.

\com{
While the other layers are important as well for 
a next generation transport services architecture,
the Flow Layer presents immediate technical challenges
that are worthy of implementation and evaluation.
This paper therefore focuses on the Flow Layer in the context of \tng;
other aspects of \tng are summarized briefly in Section~\ref{sec-arch}.
}

In this paper, we develop \tng's Flow Layer to enable PEPs in the path
to interpose on or {\em split} Flow Layer sessions,
much like traditional PEPs often split TCP sessions~\cite{rfc3135}.
Since \tng's end-to-end security and reliability functions
are implemented separately in higher layers,
this {\em flow splitting}
avoids interfering with higher end-to-end functions.
\tng's end-to-end layers treat Flow Layer sessions as ``soft state,''
and can restart a flow that fails
due to a PEP crash or network topology change,
preserving end-to-end reliability and fate-sharing.
A key technical challenge flow splitting presents
is joining the congestion control loops of consecutive path sections
to yield end-to-end congestion control over the full path,
a challenge we solve via a simple but effective technique
we call {\em queue sharing}.

\com{
@@@ move elsewhere
Other work has proposed factoring out congestion control
for other reasons,
such as congestion state sharing~\cite{
	rfc2140,balakrishnan99integrated,zhang00speeding}
and multipath transfer~\cite{al04ls-sctp}.
Flow splitting complements these ideas
and further bolsters the case for factoring congestion control
in future Internet transports.
}
Through simulations we demonstrate that flow splitting via queue sharing
can effectively address a variety of common performance issues,
such as optimizing the performance of lossy last-mile wireless links
and reducing queueing latencies on residential broadband links.
While our simulations do not attempt to analyze all relevant scenarios,
they illustrate the potential uses of flow splitting
and suggest the feasibility of implementing it via queue sharing.
We also demonstrate the feasibility of the \tng architecture
through a working user-space prototype
that functions on both real and simulated networks.
Finally, we discuss approaches to incremental deployment,
noting that with moderate costs, a \tng stack could be
(1) built entirely by rearranging existing protocols
without creating any new ones;
(2) deployed at OS level
transparently to existing applications; and
(3) made compatible with
and even benefit from existing PEPs
by using legacy TCP as an imperfect but workable ``Flow Layer.''

\com{
Previous work has decoupled congestion control from transport semantics
for other reasons,
such as congestion state sharing~\cite{
	rfc2140,balakrishnan99integrated,zhang00speeding}
and concurrent multipath transport~\cite{al04ls-sctp}.
We believe this growing list of reasons to decouple congestion control
suggests that any future Internet architecture,
whether evolutionary or ``ground-up,'' should include such a decoupling.
}

This work makes the following contributions.
First,
we identify the Internet's architectural coupling
of congestion control with end-to-end semantics in the transport layer
as the source of many of the difficulties PEPs create,
and present a clean solution based on decoupling these functions.
Second,
we introduce queue sharing as a simple but effective technique
for joining congestion control loops at PEPs in the Flow Layer.
Third,
we demonstrate that the proposed decoupling is practical and
addresses a variety of common performance issues
that concern home and business users.
\com{
Fourth,
by adding E2E-friendly flow splitting
to the list of other known reasons
for decoupling congestion control from the transport,
such as congestion state sharing~\cite{
	rfc2140,balakrishnan99integrated,zhang00speeding}
and concurrent multipath transfer~\cite{al04ls-sctp},
we further bolster the growing body of evidence
that such a decoupling in some form should have a role
in any new or evolved Internet transport architecture.
}

Section~\ref{sec-bg} of this paper
examines congestion control challenges and existing solutions.
Section~\ref{sec-arch}
briefly summarizes the \tng architecture, and
Section~\ref{sec-flow}
details flow splitting via queue sharing in the context of \tng.
Section~\ref{sec-flowsim}
uses simulations to test the feasibility and efficacy of 
flow splitting and queue sharing, and
Section~\ref{sec-impl}
describes our prototype
together with experiments confirming \tng's practicality.
Section~\ref{sec-deploy}
discusses incremental deployment strategies,
Section~\ref{sec-related}
reviews related work,
and
Section~\ref{sec-concl} concludes.


\section{The Congestion Conundrum}
\label{sec-bg}

This section first examines the origin of TCP congestion control
and the challenges it encountered as the Internet diversified,
then reviews the many approaches proposed to address these challenges
and their technical tradeoffs.

\subsection{Why is Congestion Control in TCP?}

Though network congestion was a recognized problem~\cite{
davies72control, gerla80flow},
TCP did not include congestion control
when it was first specified and deployed~\cite{rfc793}.
Only after several years of debate
about whether congestion control should be
a network or transport layer function~\cite{
	rfc896, rfc1016, finn89connectionless},
the transport layer approach took hold~\cite{
	jacobson88congestion, rfc1122}
and eventually was officially sanctioned~\cite{rfc1812}.
TCP congestion control~\cite{rfc2581}
kept routers simple and performed well on typical networks of the time.
To do so, TCP endpoints {\em infer} congestion information
from nothing but the {\em absence} of timely packet arrival,
using an implicit heuristic model of the way typical network components
are expected to behave.
But this inference approach assumes
that all devices on the path behave consistently according to this model,
an assumption somewhat contrary to the Internet's original purpose
of making {\em diverse} physical networks interoperate~\cite{clark88design},
and soon proven inaccurate~\cite{barakat99tcp}.


Arguments for end-to-end congestion control
sometimes invoke the end-to-end principle,
but the principle's original formulation~\cite{saltzer84endtoend} 
concerns {\em reliability},
and explicitly acknowledges that performance concerns
may justify in-path mechanisms augmenting (but not replacing)
end-to-end reliability checks.
The inclusion of congestion control in TCP
thus appears more a product of historical expedience
than an application of deep internetworking principles.

\subsection{Patching Up TCP Congestion Control}
\label{sec-patching}
As the Internet grew to incorporate network technologies
that violate the assumed model of network behavior underlying TCP's inferences,
a vast array of techniques appeared
to make TCP perform adequately over these new technologies.
We classify these techniques
into brute force, link-layer fixes, new inference schemes,
explicit feedback,
transport interposition, and mid-loop tuning.


{\bf Brute Force:}
A seductively easy ``sledgehammer solution'' to many TCP ills
is simply to open parallel TCP streams over one path,
either at transport~\cite{sivakumar00psockets}
or application level~\cite{allman96application\abbr{}{,lee01applied}}.
\com{ refs for data striping across a {\em single} path:
	XFTP~\cite{allman96application} (for satellite links),
	PSockets~\cite{sivakumar00psockets} (for grid environments),
	bbcp~\cite{XXX} (also for grid environments)
	- "Peer-to-Peer Computing for Secure High Performance Data Copying"
	GridFTP~\cite{lee01applied} (also for grid environments).
	why striping improves performance:~\cite{hacker02endtoend} }%
This approach effectively amplifies TCP's aggressiveness,
boosting throughput at the cost of fairness~\cite{
	floyd99promoting\abbr{}{,hacker02endtoend}}.
MulTCP~\cite{crowcroft98differentiated} achieves the same effect
in a single TCP stream.

{\bf Link-Layer Fixes:}
Most wireless networks perform link-layer retransmission
to reduce TCP's misinterpretation of radio noise as congestion,
at the costs of introducing delay variation and reordering,
and/or risking redundant retransmissions
by the two layers~\cite{wong99improving, inamura04impact}.
Forward error correction can reduce losses
while minimizing delay and reordering,
but incurs bandwidth overhead on {\em all} packets,
not just those affected~\cite{chockalingam99wireless}.
While link-layer fixes are useful,
they incur unnecessary costs to
delay/jitter-sensitive and loss-tolerant non-TCP traffic,
and cannot address other issues affecting TCP
such as high end-to-end round-trip times.

{\bf New Inference Schemes: }
\com{
Network technologies vary in many ways,
not only in physical and logical signaling, topologies, and packet formats,
but in performance characteristics.
A single communication path through a diverse ``internetwork''
may traverse links based on multiple network technologies,
but TCP's end-to-end approach to congestion control
implicitly assumes that all links behave consistently:
e.g.,
that losses indicate congestion,
total link capacity is relatively stable,
and reordering is rare.
History has proven this consistency assumption wrong,
as the introduction of each significant new networking technology
spawned its own body of research on modifying TCP congestion control
to perform adequately over that technology:
e.g.,
}
Each significant new networking technology
has spawned efforts
to modify TCP endpoints to make better congestion control inferences
when run over that technology:
e.g., for
	mobile~\cite{caceres95improving},
	satellite~\cite{akyildiz01tcp},
	wide-area wireless~\abcite{sinha02wtcp,casetti02tcp}{
		fu03tcp,cen03endtoend},
	high-speed~\abcite{
		rfc3649,kelly03scalable}{
			xu04binary,
			wang05tcp,king05tcp,
			bhandarkar06ltcp,wei06fast,tan06compound,
			xu07extending},
	and ad hoc~\cite{lochert07survey} networks.
But there is an elephant in the room:
in a diverse {\em inter}network,
one path may cross several technologies in turn---%
e.g., a wired LAN, then a satellite uplink,
a high-speed transatlantic cable,
and finally a remote ad hoc network.
But we can choose only {\em one} end-to-end scheme for any single path;
separate schemes tuned to each technology are insufficient
if none performs well on the combination.
The extensive parallel literatures
on high-speed~\cite{baiocchi07yeah}
and wireless~\cite{lochert07survey}
congestion control schemes
rarely interact or experiment over {\em diverse} paths,
giving us little optimism
that any inference-based end-to-end scheme
will perform well on {\em all} current, let alone future, network technologies.

New inference schemes also face the burden of competing
fairly with legacy flows~\cite{jin03spectrum},
a constraint that may be in
conflict with the goals of the new scheme itself.
TCP Vegas~\cite{brakmo95tcp}, for example,
works well and minimizes end-to-end delay if run alone on a network,
but cannot compete fairly with traditional TCP flows~\cite{mo99analysis},
because the signal Vegas responds to---queue build-up---is
fundamental to prevailing loss-based congestion control.
Vegas can be modified to compete fairly
by adding a loss-based component~\cite{tan06compound},
but doing so eliminates Vegas's benefit of low delay.
\com{	XXX redundant with below, but would be nice to tie them together...
The only means of employing and using these new schemes is to
isolate them from the legacy Internet, to deploy them one
administrative domain at a time.
}

\xxx{
Even where one end-to-end congestion control scheme
may be technically adequate,
the Internet's inertia
makes it difficult to agree on and deploy
new end-to-end schemes\abbr{}{~\cite{shenker94making}}.
Any new scheme encounters resistance
unless it is ``TCP-friendly''~\cite{jin03spectrum}---%
no more aggressive than TCP Reno---%
since the new scheme's flows
will compete with Reno streams ``in the wild.''
\com{TCP-friendliness severely constrains evolution,
however,
since any new scheme that wishes to improve steady-state throughput
under given (fixed) network conditions
by definition {\em must} send more data per unit time---%
i.e., be more aggressive---%
under those network conditions.
Thus, only schemes
that do not improve steady-state throughput
but only affect other metrics such as delay and/or jitter,
such as TCP Vegas~\cite{brakmo95tcp} or TFRC~\cite{
	floyd00equation\abbr{}{,rfc3448}},
tend to be politically accepted.
}
But
since the Internet does not
{\em enforce} TCP-friendliness~\cite{floyd99promoting},
selfish or unaware users
can and do deploy unfairly aggressive mechanisms anyway---%
e.g., in the form of TCP-unfair UDP flows~\cite{chung02fairplayer}
or concurrent TCP flows~\cite{liu01impact\abbr{}{,hacker02endtoend}}.
}

{\bf Explicit Feedback:}
Schemes like CSFQ~\cite{stoica98core} and XCP~\cite{katabi02internet}
for high-speed networks,
and ATCP~\cite{liu01atcp} and ATP~\cite{sundaresan03atp}
for wireless networks,
require routers
to provide more information,
such as explicit notification of
losses~\cite{balakrishnan98explicit},
congestion~\cite{rfc3168},
or link failures~\cite{holland02analysis},
to the TCP endpoints.
But Internet router upgrades are feasible today only if done incrementally,
one administrative domain at a time.
Since an end-to-end path may cross several domains,
congestion control schemes requiring router upgrades
cannot be deployed end-to-end but only in restricted domains.

{\bf Transport Layer Interposition:}
Network operators
often do not control end hosts
and have little leverage to make users adopt
new end-to-end congestion control schemes;
they must instead make prevalent TCP implementations
perform well
by managing heterogeneity {\em within} the network.
TCP-splitting PEPs~\cite{rfc3135}
interpose on transport connections
as they cross specific links or administrative boundaries,
e.g., optimizing loss-prone~\cite{yavatkar94improving}
or mobile~\cite{bakre97implementation} wireless links.
These PEPs ``split'' an end-to-end connection
into multiple sections,
applying specialized algorithms
to network segments exhibiting non-traditional behavior.
A PEP cannot interpose on the transport's congestion control loop
without interposing on its semantic functions as well,
however,
breaking TCP's end-to-end reliability and fate-sharing~\cite{clark88design}.
\abbr{}{
Unlike temporary link or router failures,
which merely require the endpoints to retransmit data
once the network heals,
a PEP failure necessarily breaks the end-to-end connection,
because ``hard'' state,
such as packets the PEP has accepted and acknowledged
but not yet successfully forwarded,
is lost.
}
Transport interposition also interferes
with end-to-end IPsec~\cite{rfc4301},
since interposition is effectively
a ``man-in-the-middle attack''~\cite{rfc3135}.

\xxx{ basic cross-layer indirection argument for mobility:
	~\cite{badrinath93handling} }

{\bf Mid-loop Tuning:}
\label{sec-bg-midloop}
An alternative to interposition
is for a PEP to manipulate a connection
from the {\em middle} of a congestion control loop;
we refer to this approach as {\em mid-loop tuning}.
For mobile/wireless networks,
Snoop~\cite{balakrishnan95improving}
caches TCP segments and retransmits them
when it detects non-congestion packet loss;
M-TCP~\cite{brown97mtcp}
manipulates TCP's receive window to trick the sender
into throttling transmission without reducing its congestion window.
PEPs for high-speed networks
use ACK splitting~\abcite{
	jin99spack,cisco04rate}{hasegawa07receiver}
to trick the sender into into increasing its congestion window more quickly,
and window stuffing~\cite{cisco04rate}
to compensate for end hosts with receive buffers
too small for the bandwidth-delay product.

While mid-loop tuning avoids violating TCP's end-to-end semantics,
it is still incompatible with IPsec,
as IPsec prevents PEPs from seeing or modifying
the relevant transport headers.
Mid-loop tuning may also interfere destructively
with modifications to end host congestion control algorithms,
as occurred between Snoop and SACK~\cite{vangala03tcp}.
Multiple PEPs residing on one end-to-end path
unbeknownst to each other can also interfere:
e.g., if a TCP connection crosses $k$ wide-area links,
each with an ACK splitting PEP
that multiplies the sender's congestion window increase rate by a factor of $n$,
the combination may unexpectedly
multiply the sender's aggressiveness by $n^k$.
\xxx{also, cite Max and Krishna's paper on traffic shaping.}
Finally, mid-loop tuning by definition exploits
a transport's vulnerability to manipulation,
and such vulnerabilities are exploitable
for malicious purposes as well;
parallel research efforts
are now devoted to closing these same vulnerabilities~\cite{
	savage99tcp,stanojevic08can}.

\com{
\subsubsection{A Patchwork of Band-aids}

\xxx{ this overlaps with the above text...}

The first response to new network technology generally comes from the
network operator in the form of a mid-loop tuning mechanism, primarily
because end-to-end schemes must be deployed by end-users, and network
operators have no control over this installed-base. It is a tempting
solution in the short-term, but has two unintended consequences.

First, having fixed the ``problem'' properties that the new technology
introduces (such as random packet loss, for instance, that diverges
from expected network properties), neither end-users nor operators see
a need to push towards the architecturally cleaner end-to-end
solutions. End-to-end congestion control solutions are unable to
attract the traction they need to get deployed extensively.

Finally, since mid-loop mechanisms typically depend on exploiting
protocol vulnerabilities~\cite{cisco04rate,hasegawa07receiver} which
are constantly being fixed~\cite{savage99tcp,stanojevic08can}, this
evolutionary process increasingly resembles an arms race---albeit with
no explicitly competing authorities---leading to a patchwork of
complex band-aids across the Internet~\cite{rfc3027,rfc3135} that aim
to fix ``problems'' introduced by new network technologies.
}

\section{Refactoring the Transport}
\label{sec-arch}


This section briefly describes \tng's overall architecture
to provide context for exploring flow splitting in the rest of the paper.
We focus on those aspects relevant
to understanding how \tng supports flow splitting,
omitting many other details of the architecture.

\subsection{Architectural Goals}

\tng's functional layering,
illustrated in Figure~\ref{fig-layers},
builds on previously proposed ideas~\cite{ford08breaking} by
decomposing the Internet's traditional transport layer
with a goal of cleanly separating
{\em network-oriented} from {\em application-oriented} functions.
We define network-oriented functions
to be those concerning
reliable and efficient network operation:
functions that network operators care about,
such as who is using the network and how it is performing.
We define application-oriented functions
as those concerning only application endpoints,
such as application content
and the end-to-end transport abstractions that applications build on.
\tng's lower Endpoint and Flow Regulation Layers implement
what we consider the network-oriented functions
of endpoint identification and congestion control, respectively,
while \tng's Isolation and Semantic Layers implement
the application-oriented functions
of end-to-end security and reliability.

\com{

(Move this discussion to end of section if necessary - jana)

One reasonable view is that
the Internet's ``narrow waist'' between IP and TCP
originally represented precisely such a boundary---%
the design of IP fragmentation~\cite{rfc791} clearly assumes
that only the endpoints need to see past the IP header---%
but the {\em de facto} boundary has effectively moved upwards~\cite{
	rosenberg08udp}
to encompass TCP, UDP, and perhaps even HTTP
as higher-level protocol functions
came to be of concern to network operators
and their ever smarter middleboxes.

XXX note somewhere (discussion section at end?)
that functions like data deduplication and data-oriented transport (DOT)
might eventually constitute even higher-level ``network-oriented functions.''

Regardless of perspective,

}

We acknowledge that the ``correct'' boundary
between network-oriented and application-oriented functions
is not clear-cut and may be a moving target.
\tng's contribution as an architecture
is not to find a perfect or complete decomposition
of the transport layer,
but to identify specific transport functions
that have proven in practice to be ``network-oriented''
contrary to their traditional placement in the transport layer,
and to construct a new but incrementally deployable layering
that reflects this reality
and restores the ``end-to-endness''
of the remaining application-oriented functions.

The following sections briefly outline each \tng layer.

\subsection{The Endpoint Layer}
\label{sec-endpoint}

As in the OSI model~\cite{zimmermann80osi},
TCP/IP breaks application endpoint identifiers
into Network Layer (IP address)
and Transport Layer (port number) components,
including only the former in the IP header
on the assumption that the network need know
only how to route to a given host,
and leaving port numbers to be parsed and demultiplexed by the transport.
As the Internet's size and diversity exploded, however,
network operators needed to 
enforce access policies
that depend on exactly {\em who} is communicating---%
not just which hosts, but which applications and users.
Now-ubiquitous middleboxes such as  
Firewalls~\cite{rfc2979},
traffic shapers~\cite{ferrill06network},
and NATs~\cite{rfc3022}
must therefore understand transport headers
in order to enforce these network policies.
Since middleboxes cannot forward traffic for transports
whose headers they do not understand,
new transports have become effectively undeployable
other than atop TCP or UDP~\cite{rosenberg08udp}.

Recognizing that communicating rich endpoint information
is a network-oriented function relevant to in-network policy enforcement,
\tng factors this function into its Endpoint Layer
so that middleboxes can extract this information
without having to understand application-oriented headers.
\tng reinterprets UDP~\cite{rfc768}
as an initial Endpoint Layer protocol
already supported by most middleboxes,
but we are evolving \tng to incorporate ideas on
richer endpoint identities~\abcite{
	touch06tcp}{freire08tcp},
NAT traversal~\cite{
	ford05p2p,guha05characterization,biggadike05natblaster},
middlebox signaling~\abcite{
	upnp01igd,cheshire05nat}{rfc5190,stiemerling08nat,
				woodyatt08application},
NAT-friendly routing~\cite{
	walfish04middleboxes,guha04nutss},
and other related ideas outside the scope of this paper.

\subsection{The Flow Regulation Layer}
\label{sec-flow-summary}

As \tng's Endpoint Layer factors out endpoint identification,
the Flow Regulation Layer similarly factors out
performance related functions such as congestion control,
with the recognition that these functions
have likewise become ``network-oriented'' in practice
as discussed in Section~\ref{sec-bg}.
The Flow Layer assumes that the underlying Endpoint Layer
provides only best-effort packet delivery between application endpoints,
and builds a {\em flow-regulated} best-effort delivery service
for higher layers to build on.
In particular,
the Flow Layer's interface to higher layers
includes an explicit signal indicating when the higher layer
may transmit new packets.

To perform this flow regulation,
the Flow Layer may either implement
standard TCP-like congestion control~\abcite{
	jacobson88congestion}{rfc2581},
or,
as we discuss in later sections,
may use more specific knowledge
of an underlying network technology or administrative domain.
In the longer term,
we envision \tng's flow layer incorporating
additional performance-related mechanisms
such as end-to-end multihoming~\cite{rfc4960},
multipath transmission~\cite{
  magalhaes01transport
  \abbr{}{,hsieh02ptcp,zhang04transport,al04ls-sctp,iyengar06concurrent}},
and forward error
correction\abbr{}{~\cite{banerjea96simulation,nguyen03path}}.

\com{
Deploying new network technologies
or new congestion control schemes
in the Internet faces significant challenges today;
and doing so without adversely impacting the performance
of existing end-to-end flows is a very difficult,
if not impossible,
proposition.
Factoring congestion control out of the transport 
is necessary for better adoption of both 
new network technologies and new congestion control schemes.
Separating flow performance related concerns has significant other advantages:
}

\subsection{The Isolation Layer}
\label{sec-iso}

Having factored out network-oriented transport functions
into the Endpoint and Flow Layers,
the optional Isolation Layer 
``isolates'' the application from the network,
and protects the ``end-to-endness'' of higher layers.
This isolation includes two elements.
First, the Isolation Layer
protects the application's end-to-end communication
from interference or eavesdropping within the path,
via transport-neutral cryptographic security as in IPsec~\cite{rfc4301}.
Second, the Isolation Layer
protects the application and end-to-end transport
from unnecessary exposure to details of network topology and attachment points,
by implementing location-independent endpoint identities
as in HIP~\cite{rfc4423} or UIA~\cite{ford06persistent},
which remain stable
even as devices move or the network reconfigures.
The Isolation Layer's interface to higher layers
is functionally equivalent to the interface exported by the Flow Layer,
but with transformed packet payloads and/or endpoint identities.

We believe the Isolation Layer
represents a suitable location for end-to-end security
precisely because it defines the boundary between network-oriented
and application-oriented functions,
thus ensuring integrity and security of the latter,
while allowing middleboxes to interact with the former.
In contrast with SSL/TLS~\cite{rfc4346},
the Isolation layer is neutral to transport semantics
and does not need to be adapted to each transport~\cite{rfc4347}.
In contrast with IPsec's standard location immediately above IP,
the Isolation Layer does give up the ability to protect
Endpoint and Flow Layer mechanisms from off-path DoS attacks
as IPsec protects TCP's signaling mechanisms,
but if standard non-cryptographic defenses against such attacks~\cite{
	rfc1948,rfc4987} are deemed insufficient,
then IPsec authentication can still be deployed in \tng
underneath the flow layer,
ideally via a delegation-friendly scheme~\cite{walfish04middleboxes,guha04nutss}
permitting controlled interposition by middleboxes.


\subsection{The Semantic Layer}
\label{sec-sem}

\tng's Semantic Layer
implements the remaining application-oriented end-to-end transport functions,
particularly end-to-end reliability.
In the case of TCP,
these functions are
all those in the original TCP protocol~\cite{rfc793}
except port numbers,
including acknowledgment and retransmission,
order preservation,
and receive window management.
\com{
and include:
\begin{itemize}
\item	{\bf Acknowledgment and retransmission:}
	maintaining the end-to-end reliability
	of TCP's byte stream abstraction.
\item	{\bf Ordering:}
	delivering bytes in the order they were sent.
\item	{\bf Urgent data:}
	indicating out-of-band signaling in the stream.
\item	{\bf Flow control:}
        an end-to-end, application-oriented function
	limiting the sender's transmission
	to the rate at which the receiver can accept data.
\end{itemize}
}
Other application-visible semantics,
such as RDP's reliable datagrams~\cite{rfc1151}
and SCTP's message-based multi-streaming~\cite{rfc4960},
could fit equally well into \tng's Semantic Layer
as distinct protocols.

The Semantic Layer's interface to lower layers
differs from that of traditional Internet transports in two ways.
First, a \tng semantic protocol
uses the Endpoint Layer's endpoint identities
(possibly transformed by the Isolation Layer)
instead of implementing its own port number demultiplexing.
Second, a \tng semantic protocol implements no congestion control
but relies on the underlying Flow Layer to signal
when packets may be transmitted.
The Semantic Layer's interface to higher layers (e.g., the application)
depends on the transport semantics it implements,
but need not differ in any application-visible way
from existing transport APIs---%
a fact that could aid deployment as we discuss later
in Section~\ref{sec-deploy}.

\section{Flow Splitting in \TNG}
\label{sec-flow}


With the architectural context in place, 
we now focus on \tng's support for flow splitting at the Flow Regulation Layer,
in order to support in-path congestion control specialization
without interfering with end-to-end transport functions.
\com{
Flow Layer for short---
that factors out congestion control from the Transport Layer.
We now clearly define its purpose and scope,
explore its main benefit of enabling middleboxes
to interpose on congestion controlled paths
without breaking end-to-end semantics,
and describe joining flow segments through queue sharing.
}

\subsection{Flow Middleboxes}
\label{sec-flow-def}

\begin{figure}[tbp]
\centering
\includegraphics[width=0.48\textwidth]{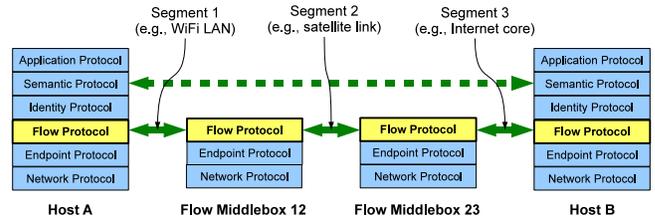}
\caption{An end-to-end path composed of multiple Flow Layer segments.
	Flow middleboxes can optimize network performance
	based on the properties
	of a specific segment, such as a satellite link.}
\label{fig-flow}
\end{figure}

\tng enables network operators to specialize congestion control
and other flow performance concerns
by deploying devices we call {\em flow middleboxes}
at network technology and administrative boundaries.
As illustrated in Figure~\ref{fig-flow},
a flow middlebox interposes on a Flow Layer session,
effectively terminating one congestion control loop
and starting another for the next section of the path.
Each section may consist of one or many Network Layer hops:
flow splitting does not imply
hop-by-hop congestion control~\cite{mishra92hopbyhop\abbr{}{,yi07hopbyhop}},
although the latter might be viewed as a limit case of flow splitting.

Each flow section may use any congestion control scheme
operating according to standard principles;
the key technical challenge is joining
these independent segments
to form a single flow providing end-to-end congestion control
to higher layers,
a challenge we address in Section~\ref{sec-impl-flow}.

While flow middleboxes are similar to PEPs,
they avoid the problems of PEPs
discussed in Section ~\ref{sec-patching}.
Since \tng's Flow Layer implements 
only performance-related functions,
Flow middleboxes
interpose on only these functions
without interfering with end-to-end functions.
Flow middleboxes maintain
only performance-related ``soft state;''
end-to-end functions can recover from
a flow middlebox failure
since reliability and connection-related ``hard state''
are located at the endpoints.
We demonstrate this fate-sharing in \tng
through experiments using our prototype implementation
in Section~\ref{sec-impl-fate}.

\com{
Flow splitting also avoids the dangers of mid-loop tuning
discussed in Section~\ref{sec-patching}
by operating at a 
transport-neutral
network-oriented 
layer in the stack,
and by enabling explicit negotiation 
with other flow middleboxes on the end-to-end path.
}
\subsection{Uses of Flow Splitting}
\label{sec-seg}

\com{
performance-oriented middleboxes to exist. We not only recognize their
inevitability, but also believe that these interposers have immense
value in the evolution of the Internet infrastructure by allowing link
and network technologies to evolve without the baggage of preconceived
notions about their characteristics.

flow happens between flow terminators, but the logical end-to-end
session at the transport layer is oblivious of these interposers.
Flow terminators do not concern themselves with reliability but with flow
rate, error recovery, congestion control and performance-enhancement
strategies for the subsequent path section. They may choose to
explicitly interpose on the data path and thus rewrite the Regulation
header, or they may choose to remain transparent as is the current
case with most middleboxes. In any case, the endpoint identities are
preserved at this layer. Note that in our conception of the Regulation
layer, the flow terminators are {\em not} required to be on both the
forward and reverse paths between the two endpoints.
}

Flow splitting can be used to improve communication performance
in at least three ways, which we summarize here:
reducing per-section RTT,
specializing to network technology,
and administrative isolation.

\xxx{discuss rewriting vs non-rewriting interposers}
\xxx{clarify how it works if terminators are not on both fw \& reverse paths.}

\xxx{Could flow splitting be used to address the Incast problem?
	"Measurement and Analysis of TCP Throughput Collapse
	in Cluster-based Storage Systems"}

{\bf Reducing Per-Section RTT:}
A TCP flow's throughput is adversely affected by large round-trip time (RTT),
especially in competition with flows of smaller RTT~\cite{
	floyd91connections\abbr{}{,mathis97macroscopic,lakshman97performance}}.
Further,
since information takes one RTT
to propagate around the control loop,
{\em any} end-to-end scheme's responsiveness
to changing conditions
is limited by RTT.
Subdividing a path into shorter sections
reduces each section's RTT to a fraction of the path's RTT,
which can improve both throughput and responsiveness.
Proponents of 
hop-by-hop congestion control schemes
for packet-switched~\cite{mishra92hopbyhop\abbr{}{,kortebi04crossprotect}},
cell-switched~\cite{kung93fcvc\abbr{}{,ozveren94reliable}},
and wireless networks~\cite{yi07hopbyhop\abbr{}{,scofield07hopbyhop}}
have noted this benefit.
The Logistical Session Layer~\cite{swany04improving}
similarly leverages the reduced RTT of split paths
to improve wide-area grid performance.

\xxx{CDNs?}
\xxx{pipelining?}

{\bf Specializing to Network Technology:}
The literature reviewed in Section~\ref{sec-bg}
amply demonstrates that
the best congestion control scheme for a communication path
often depends on underlying network characteristics.
\xxx{ ~\cite{barakat00tcp}. 
	(already said this above, but maybe integrate...)
Classic TCP congestion control~\abcite{jacobson88congestion}{rfc2581}
performs well on wired LANs and the Internet core,
but poorly on networks that are loss-prone due to
transmission errors\abbr{}{~\cite{desimone93throughput}}
or mobility\abbr{}{~\cite{caceres95improving}},
and on long-delay connections such as satellite links\abbr{}{~\cite{
	allman97tcp\abbr{}{,partridge97tcpip,lakshman97performance}}}
or wireless wide-area networks\abbr{}{~\cite{rfc2757}}.
}
\com{
	formal analysis of TCP sensitivity to high BDP and random loss,
	unfairness to high-RTT flows:~\cite{lakshman97performance}
	more experimental:~\cite{mathis97macroscopic}
}
Flow middleboxes deployed at the boundaries of a network domain
can implement a congestion control specialized to that domain,
taking advantage of a more precise knowledge
of the domain's characteristics
from which to make inferences,
and/or leveraging explicit feedback mechanisms~\cite{
	stoica98core,katabi02internet,
	balakrishnan98explicit,rfc3168,holland02analysis}
supported only within that domain.
Although one path may traverse many such boundaries,
each middlebox need only understand the properties
of the adjacent path sections,
reducing the ``end-to-end'' challenge of managing flow performance
across an arbitrary set of network technologies
to the more tractable challenge
of interfacing technologies in pairwise combinations.
The fact that one ``side'' of each flow middlebox
is usually a standard wired LAN
simplifies the challenge further.

\com{
Since integrating diverse networks
is a fundamental goal of the Internet~\cite{clark88design},
we must assume that
any communication path may traverse several network types,
each of which might place conflicting requirements
on any single end-to-end congestion control scheme.
New end-to-end schemes are available
for high-bandwidth, long-delay links~\cite{
	rfc3649
	\abbr{}{,kelly03scalable,wang05tcp,xu04binary,wei06fast,xu07extending}},
and others for mobile ad hoc networks~\cite{lochert07survey},
but will any one scheme perform well on a path
that includes links of both types (and others)?
}
%

\xxx{
link-layer retransmit plus FEC:~\cite{paul95asymmetric}
}

\xxx{
ACK splitting, window stuffing, drop reporting?	(RBSCP) \cite{cisco04rate}
}

{\bf Administrative Isolation:}
\label{sec-seg-admin}
Flow splitting 
enables administrators to split a Flow Layer path
at domain boundaries
and deploy a new congestion control scheme within the domain
under controlled conditions,
while maintaining TCP-friendliness on other sections
of paths crossing the domain.
Even for legacy flows not conforming to \tng's model---%
e.g., flows with congestion control embedded in the Transport Layer
or no congestion control at all---%
administrators can enforce
the use of a particular congestion control scheme within a domain
by encapsulating legacy streams in a Flow Layer ``tunnel'' as a
mechanism using per-flow state at border routers/flow middleboxes
to deploy new congestion control schemes within a domain~\cite{
	stoica98core\abbr{}{,katabi02internet}},
or to enforce TCP-friendliness~\cite{
	rangarajan99eruf\abbr{}{,albuquerque04network}}
or differential service agreements~\cite{
	habib01unresponsive\abbr{}{,wu01direct}}.
Flow splitting thus gives administrators the freedom to choose
schemes like Vegas~\cite{brakmo95tcp}
for their desirable properties,
while isolating the chosen scheme from competition with legacy Reno flows
and avoiding the yoke of TCP-friendliness.

\com{ fairness enforcement mechanisms:
	network-based:
	- Floyd, "Promoting ...", 1999
	- Flow-Aware Networking
	border-based:
	- Rangarajan, "ERUF: early regulation of unresponsive [flows]", 1999
	- Core Stateless Fair Queueing, XCP?
	- Network Border Patrol, 2000/2004
	border-based diffsrv:
	- Habib, "Unresponsive flow detection ... using diffsrv", 2001
	- Wu, "Direct Congestion Control Scheme ... for diffsrv", 2001
	...
}

\subsection{Joining Flow Sections}
\label{sec-impl-flow}

As mentioned earlier,
the primary technical challenge in implementing flow splitting
is joining multiple independently congestion controlled sections
to form an end-to-end congestion controlled path.
Existing TCP splitting PEPs leverage
the buffer management and receive window control
that TCP's reliable byte stream abstraction provides,
but these heavyweight abstractions are not well suited
to \tng's best-effort, packet-oriented Flow Layer.

\tng addresses this challenge
through a simple technique we call {\em queue sharing}.
We assume each flow middlebox along a split path has a queue
in which it holds packets it has received on one section
but not yet forwarded on to the next section.
With queue sharing,
the middlebox treats this queue as the meeting point for the two sections,
with each section's congestion control loop
taking a role in the queue's management:
the two adjacent sections thus ``share'' this queue.

\begin{figure}[t]
\centering
\includegraphics[width=0.45\textwidth]{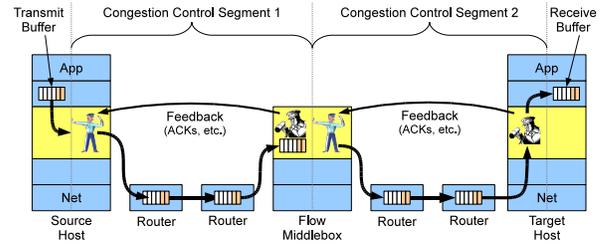}
\caption{Joining Sections through Queue Sharing}
\label{fig-joinqueue}
\end{figure}

Consider for example data sent from the source host across Section 1
and arriving at the flow middlebox in Figure~\ref{fig-joinqueue}.
Instead of acknowledging a data segment immediately upon reception
as TCP would,
the flow middlebox {\em silently} deposits the packet in its shared queue.
The transmit side of the middlebox's congestion control logic for Section 2,
meanwhile,
determines when the middlebox may remove packets from the shared queue
and transmit them over Section 2 to the target host.
When Section 2's congestion control logic
decides a packet may be transmitted,
the middlebox removes and transmits a packet from the shared queue,
and {\em only then}
allows the receive-side logic for Section 1
to acknowledge the packet's receipt.
The middlebox in effect treats the shared queue
as if it were the last router in Section 1,
including the queue in Section 1's congestion control loop
so that the sender on Section 1 (the source host in this case)
throttles its transmit rate if this
or any other Section 1 router queue fills.

Suppose the path's bottleneck
is one of the routers in Section 2.
As the bottleneck router's queue fills,
Section 2's congestion control scheme detects this bottleneck,
typically by sensing either a packet loss or delay increase
depending on the congestion control scheme.
The flow middlebox in response cuts its transmission rate over Section 2,
thereby decreasing the rate at which it removes packets from the shared queue.
As the shared queue fills,
Section 1's transmitter---the source host---notices
either a loss or a delay increase and cuts its transmission rate in turn.

Queue sharing is simple
and works with any congestion control algorithm
as long as the middlebox manages the shared queue
in the proper fashion for routers in the section feeding the queue.
If that section consists of standard Internet routers,
then the shared queue may be a standard drop-tail queue,
or a RED~\cite{floyd93random}
or ECN-marking~\cite{rfc3168} queue to improve performance.
If the feeding section uses XCP~\cite{katabi02internet},
then the shared queue must behave like an XCP router,
tagging packets flowing through it with congestion information.

\abbr{}{
The shared queue technique has one clear disadvantage:
if most or all sections in a path use traditional loss-based congestion control
and the congestion bottleneck is in the last section,
then the shared queue in {\em each} middlebox on the path must fill
before the original sender finally cuts its transmission rate,
resulting in all of these full queues adding to the end-to-end delay.
Keeping the number of sections small
and/or using delay-sensitive congestion control on some of them
may mitigate this delay-increasing effect,
but eliminating it entirely may require
another section joining technique.

\paragraph{Congestion Control Layering}

Congestion control layering takes a more explicitly hierarchical approach
to joining flow sections into an end-to-end path:
each section runs its own fully-independent congestion control loop
that shares no router queues with adjacent sections,
but instead we treat each of these congestion-controlled section
as a single ``virtual network link'' in a higher-level or ``overlay'' path,
and we run a second congestion control scheme on that overlay path
to provide end-to-end congestion control. The choice of the overlay congestion control scheme
used to join the individual sections
is likely to be critical to performance, and while we have not explored this design yet, we think it holds promise.

\com{The choice of the overlay congestion control scheme
used to join the individual sections
is likely to be critical to performance,
...
XXX XDR, but be brief...
}
}

\subsection{Limitations of Queue Sharing}
\label{sec-flow-limits}

Queue sharing is appealing due to its simplicity
and practical applicability as explored in following sections,
but it has at least two limitations
that may suggest future refinements or alternative flow joining techniques.

First, queue sharing assumes that
the middlebox maintains a separate queue per flow,
which may be expensive in middleboxes supporting many flows.
This situation is still an improvement
over the per-flow state requirements of TCP splitting PEPs, however,
which typically need {\em two} queues in each direction---%
a receive buffer for the previous TCP session
and a transmit buffer for the next.

Second, since queue sharing essentially
transforms a downstream section's congestion
into ``backpressure'' on upstream middleboxes' shared queues,
congestion-related overheads
can accumulate across these queues.
If all sections of a path
use loss-based congestion control~\cite{rfc2581},
for example,
and the last section contains the bottleneck,
then not only the bottleneck router queue
but each upstream middlebox queue fills
before this backpressure reaches the sending endpoint,
exacerbating the loss-based scheme's delay-inducing effects.

\com{
Third, queue sharing assumes that adjacent path sections
at least operate on generally compatible, ``online'' timescales.
If path sections are so mismatched
that a downstream bottleneck section admits packets so slowly
that \ldots XXX ???
}

A possible alternative to queue sharing
is to layer one end-to-end congestion control loop
atop a series of per-section control loops.
The Flow Layer might use
XCP~\cite{katabi02internet} end-to-end, for example,
treating the lower-level per-section congestion control loops
as ``virtual links''
as seen by the upper-level XCP control loop.
Such an approach might address the above issues,
at the cost of requiring greater end-to-end coordination;
we leave such alternatives to future work.

\com{
Flow aggregation can thus implement policies
similar to those motivating hierarchical fair queuing schemes~\cite{
	bennett96hierarchical\abbr{}{,hogan96hierarchical}},
without changing interior routers.
}

\xxx{	add this back in some form, when we have time

While the Flow Layer's primary motivation
is to meet what we view as an urgent need for an architecture
that cleanly supports flow splitting,
in the long term we see the Flow Layer as a useful architectural layer
in which to implement more general and ambitious enhancements
to flow performance properties.
We outline two such categories of enhancements here:
multipath communication and flow aggregation.

\xxx{ this section needs to be reworked and shortened.}

\subsubsection{Multipath Communication}

\com{
There are many ways to exploit alternate network paths
via {\em multipath communication},
e.g., in order to 
which can operate in various network layers:
e.g., link aggregation~\cite{ieee802-3},
multipath routing~\cite{lee01split,marina01ondemand},
traffic dispersion~\cite{gustafsson97literature\abbr{}{,maxemchuk75dispersity}},
transport multihoming~\cite{
	rfc4960,hsieh02ptcp,iyengar06concurrent,	
	magalhaes01transport},				
and multiple description streaming~\cite{
	apostolopoulos02multiple} in the Application Layer.
}

There are many ways to exploit alternative network paths
to improve reliability~\abcite{
	gustafsson97literature}{
        park97highly,
	maxemchuk75dispersity,					
	raju99new,						
	rfc4960},						
balance load~\abcite{murthy96congestion}{				
	lee01split,						
	magalhaes01transport,hsieh02ptcp,iyengar06concurrent},	
or enhance security~\cite{lou01multipath}.
To be deployable, however,
a multipath scheme
must be compatible with upper layer protocols
designed assuming single-path routing,
and must remain interoperable with single-path routing domains.
Our architecture addresses these deployment issues
by permitting end hosts and flow middleboxes
to implement multipath communication
end-to-end or in the network, 
as shown in Figure~\ref{fig-multipath}.

\com{
Our Flow Layer provides a clean framework in which to implement
{\em multipath communication}:
taking advantage of alternate network paths
to improve reliability~\cite{
	gustafsson97literature\abbr{}{,maxemchuk75dispersity},	
	park97highly,\abbr{}{raju99new,}			
	rfc4960},						
balance load~\cite{
	murthy96congestion,					
	lee01split,						
	magalhaes01transport,hsieh02ptcp,iyengar06concurrent},	
or enhance security~\cite{lou01multipath}.
}

\begin{figure}[tbp]
\centering
\includegraphics[width=0.47\textwidth]{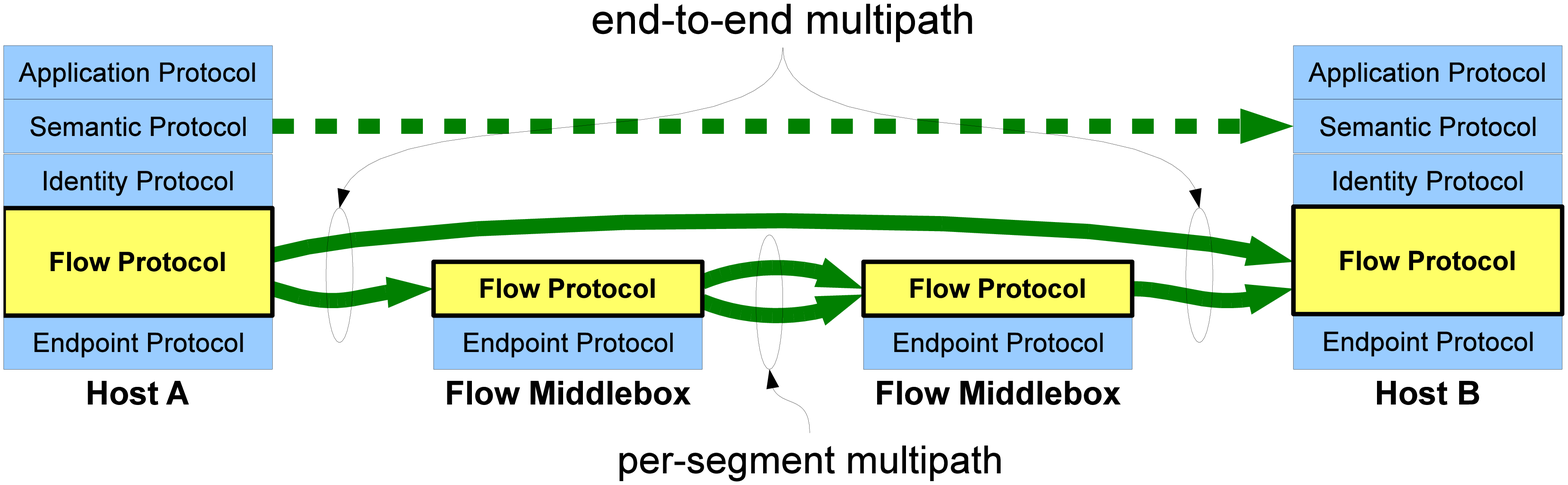}
\caption{Flow Layer multipath communication example.
	The multihomed hosts use two end-to-end paths,
	one passing through a pair of middleboxes
	implementing an in-network multipath section.}
\label{fig-multipath}
\end{figure}

\com{
Multipath communication presents two key technical challenges,
{\em providing} multipath delivery services to upper layers
without 
{\em identifying} available paths, and
{\em utilizing} them efficiently.
The following sections address each in turn,
in the context of our architecture.
}

\paragraph{Flow Layer Multihoming}

The Flow Layer provides a clean place to implement
end-to-end {\em multihoming}:
binding several endpoints together
to provide multiple paths
over the existing routing infrastructure.
In contrast with transport multihoming~\abcite{
	magalhaes01transport,rfc4960}{
	hsieh02ptcp,zhang04transport,
	argyriou03bandwidth, al04ls-sctp,
	iyengar04concurrent, iyengar06concurrent},
multihoming in the Flow Layer can benefit any transport
without interfering with transport semantics.
An address rewriting mechanism similar to shim6~\cite{shim6}
in the Flow Layer
can make all of a host's endpoints appear as one to these transports.

Flow splitting in our architecture
can also facilitate the incremental deployment of multipath routing.
A multipath routing protocol may be deployed within an administrative domain,
surrounded by flow middleboxes that can exploit available paths
in flow segments crossing that domain,
without affecting external segments (see Figure~\ref{fig-multipath}).
Alternatively, or simultaneously,
a multi-site organization might deploy flow middleboxes at site boundaries
to distribute inter-site traffic across redundant wide-area links.

\paragraph{Coping with Path Diversity in Upper Layers}

Na\"{\i}vely distributing packets among multiple paths with varying delay, 
whether end-to-end or in-network,
can confuse the congestion control and reliability mechanisms
of existing transports~\abcite{blanton02making}{
	lee02improving,zhang03rrtcp,bohacek03tcppr,	
	lim03tcp					
	}.
In our architecture,
a multihomed Flow Layer can avoid this confusion 
by implementing per-path congestion control,	
but the Transport Layer remains responsible for retransmission
and thus vulnerable to similar confusion.
To support arbitrary transports, therefore,
a multihomed Flow Layer needs to preserve
the illusion of single-path delivery,
either by using only one path at once as SCTP does~\cite{rfc4960},
or through order-preserving traffic dispersion~\cite{
gustafsson97literature}.

Multipath-aware transports~\cite{iyengar06concurrent}
and applications~\cite{apostolopoulos02multiple}
can benefit from the ability to maintain per-path state
and explicitly associate packets with paths.
Through a simple path indexing mechanism
inspired by {\em path splicing}~\cite{motiwala08path},
which we do not elaborate here for space reasons,
a multipath Flow Layer in our architecture
can expose alternative paths to upper layer protocols capable of using them,
while retaining compatibility with multipath-oblivious protocols.

\com{
 however, can avoid confusion and
control the distribution of packets at the Flow Layer when end-to-end
multipath is used. Our Flow Layer can further provide hooks that
enables a multipath-aware transport to distribute load through the
network, even on in-network multipath sections, using a simple and
weak form of source-routing as in
path-splicing\footnote{We cannot elaborate on
  this idea due to lack of space, but simply extending the
  path-splicing idea by having upper layers generate and embed in the
  packet a random nonce that dictates the path used at an in-network
  forking point allows for load balancing at the Flow and Network
  layers, with no side-effects at the Transport Layer.}~\cite{motiwala08path}.
Our architecture thus provides an evolutionary path for integrated
Transport Layer protocols that can maximally benefit from multipath
communication.
}

\com{
Since the Network Layer is responsible
for finding {\em some} route to a given host,
it is well-placed to find multiple routes to that host if available.
Many multipath routing schemes have been proposed,
for packet-switched~\cite{murthy96congestion},
circuit-switched~\cite{bahk92dynamic},
and wireless networks~\cite{lee01split}.	
The challenges of deploying such schemes incrementally on existing networks
have hindered their widespread use, however.

Flow splitting in our architecture
can facilitate the incremental deployment of multipath routing.
A multipath routing protocol may be deployed within an administrative domain,
surrounded by flow middleboxes that can exploit available paths
in flow segments crossing that domain,
without affecting external segments (see Figure~\ref{fig-multipath}).
Alternatively, or simultaneously,
a multi-site organization might deploy flow middleboxes at site boundaries
to distribute inter-site traffic across redundant wide-area links.
}
\com{
If the transport and application protocols are multipath-oblivious,
then the Flow Layer must provide serialized delivery to the Transport Layer,
but it can still transparently exploit multiple paths within the network
through flow splitting,
flow middleboxes at the borders of a routing domain
can transparently ``fork'' and ``join'' a communication path at these borders
to exploit multipath routing capability within that routing domain,
as illustrated in Figure~\ref{fig-XXX}.
If the Network Layer protocol in the multipath routing domain
supports indexed multipath delivery,
then the Flow Layer protocol on that segment
can use this delivery mode to maintain per-path congestion control state,
preventing confusion due to varying path lengths.
}

\com{

\subsubsection{Identifying Available Paths}

In {\em multipath routing}, discussed in this section,
the Network Layer is responsible for identifying multiple paths
between a single pair of endpoints,
although higher layers may be involved
in utilizing those paths efficiently.
In {\em multihoming}, discussed in Section~\ref{sec-multihome},
the Flow Layer itself identifies alternate paths,
using multiple pairs of physical endpoint addresses
to represent a single logical endpoint pair.


\subsubsection{Incremental Deployment}

\subsection{Multihoming}
\label{sec-multihome}

\paragraph{Multihoming:}
An alternative approach is to leave the Network Layer as-is,
using only one path at a time between a given pair of endpoints,
but instead attach a given ``logical'' endpoint (e.g., a host)
at multiple ``physical'' endpoint addresses
with different network attachment points.

}

\subsubsection{Flow Aggregation}

Finally,
the Flow Layer provides a clean point
at which to {\em aggregate} related flows when desired,
so that the intervening network treats the aggregate as one flow
(see Figure~\ref{fig-aggr}).
Flow aggregation can provide several benefits including
reuse of congestion control state
and improved fairness.
\com{	also reduced per-flow state in the network,
	but that's weak against the traditional attitude
	that the network should have *no* per-flow state... }

\begin{figure}[tbp]
\centering
\includegraphics[width=0.47\textwidth]{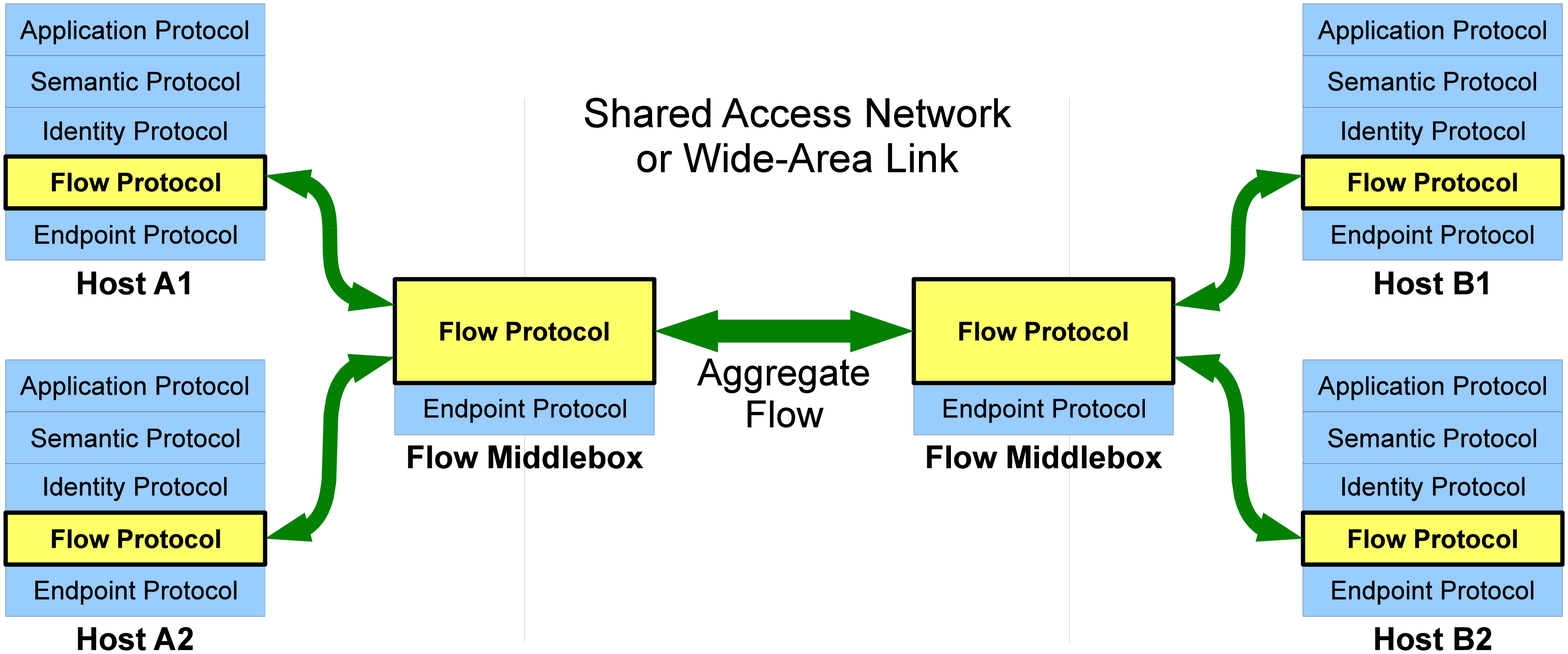}
\caption{Flow Layer aggregation example containing two end-to-end flows,
	which appear as one flow to the intermediate network.}
\label{fig-aggr}
\end{figure}

\paragraph{Reuse of Congestion Control State}

Since an aggregate of many transport instances
is typically longer-lived and represents more traffic
than any of its constituents,
measurements of the aggregate's characteristics
can benefit from a longer history and more samples.
Transport extensions have been proposed
to aggregate congestion control state
across reincarnations of one transport session~\cite{rfc1644},
across concurrent sessions~\abcite{rfc2140}{
	balakrishnan98tcp,padmanabhan98addressing,
	padmanabhan99coordinating,	
	eggert00effects},
across transport protocols~\cite{balakrishnan99integrated},
and across hosts in an edge network~\cite{zhang00speeding}.

Placing optimizations such as these in the Flow Layer
allows arbitrary transports to benefit from them,
and permits aggregation to be performed cleanly within the network
as well as end-to-end.
In our architecture, for example,
a flow middlebox
can aggregate congestion control state across the hosts in an edge network
and use that information to optimize flows crossing that middlebox
transparently,
without requiring end host modifications as in TCP/SPAND~\cite{zhang00speeding}.


\com{
Aggregation of flows from and to co-located endpoints improves
accuracy of measurement by sharing observations about network state,
and has been explored at several levels of granularity: aggregating
all TCP flows~\cite{rfc2140, balakrishnan99integrated,
  eggert00effects, padmanabhan98addressing}, aggregating all
application flows at an endhost~\cite{balakrishnan99integrated}, and
aggregating all TCP flows from within an
organization~\cite{seshan97spand, zhang00speeding}. Flow aggregation
allows flows to share and learn network congestion state from each
other instead of competing with each other for bottleneck resources,
resulting in better utilization of network resources. When aggregated,
flows provide more samples for passive measurements at the aggregation
points, thus leading to better predictions.
}

\paragraph{Fairness Control}

TCP's per-stream ``fairness'' notion
often fails to match the expectations
of users and network operators~\cite{briscoe07flow};
Flow Layer aggregation may be useful
to implement higher-level fairness policies.
For example,
an ISP may want each {\em customer}
to get equal bandwidth at bottlenecks in its network,
regardless of whether a customer
uses few transport instances
(web browsing, SSH\abbr{}{~cite{openssh}})
or many
(BitTorrent\abbr{}{~\cite{cohen03incentives}}).
To implement such a policy,
the ISP could deploy flow middleboxes at its borders
that aggregate all segments crossing its network
into one ``macro-flow'':
since each macro-flow has one congestion control context,
each macro-flow gets an equal share of congestion bottleneck bandwidth.
Most such macro-flows will connect
one customer's access router to one of
a few upstream network attachment points,
so this meta-flow fairness should approximate
a per-customer fairness policy.
Flow aggregation can thus implement policies
similar to those motivating hierarchical fair queuing schemes~\cite{
	bennett96hierarchical\abbr{}{,hogan96hierarchical}},
without changing interior routers.

\com{
A web browser may for example wish to open many transport instances at once
for  to allow AJAX applications
to open several persistent transport connections
at once~\cite{openajax-two-http},
but ensure that a web page with many persistent connections

many concurrent transport instances
to download 
An ISP or wide-area network provider may for example

wish to give each {\em customer} an equal portion
that each {\em customer} gets an equal portion
of the shared access link's bandwidth,

An application might want each of its TCP streams
to receive an equal share of bandwidth through a congestion bottleneck,
but a {\em user}
may want each {\em application} to get an equal share---%
so that the aggregate of all 20 streams
in his BitTorrent~\cite{cohen03incentives} download
get a total share equal to that of his SSH~\cite{openssh} session,
for example.
An ISP may similarly want each {\em customer}
with a particular service plan
to get an equal share of bottleneck bandwidth,
independent of how many individual hosts, users, applications, or streams
are sharing 
}

\xxx{also, flow aggregation can be a tool
to implement more useful fairness policies.}

\xxx{also, simple state reduction within the network}

\com{	I don't understand what this para is trying to say -
	rephrase to make understandable (e.g., with an example),
	or just drop?  -baf
Our Regulation Layer enables this aggregation to happen at any desired
level of granularity. A flow terminator can aggregate and share
congestion state among flows that match a desired set of filters at
any point in the end-to-end path between the application endpoints.
}

\xxx{try to explain at least briefly how it's done}

\xxx{related to aggregation and fairness:
	Nandy, "Aggregate flow control", INFOCOM 2001
	Mahajan, "Controlling high bandwidth aggregates", 2002
	Xiang, " Regulating best effort flows on per domain basis", 2003
}

\com{	XXX
\subsection{Other performance enhancements}

The flow layer can carry application preferences as soft QoS requests,
which interposing flow terminators propagate further.  Based on these
preferences and link/segment characteristics, interposing flow
endpoints can negotiate and choose appropriate congestion control
schemes for flows (or flow aggregates).

One consequence of this architecture is that there are two overlapping
feedback loops: one loop at the Regulation Layer operating at the
scale of path sections, and one at the Transport Layer operating
end-to-end. When these loops overlap fully, redundancy can be avoided
by piggy-backing both layers' information in the same ``ack''. In the
general case however, there may be redundancy in ack information on
the return path. We think that this redundancy is a small price to pay
for the clean separation of concerns and evolvability achieved.

\xxx{clarify what this means in terms of existing transports:
e.g., which transports overload acks this way, which don't.
Also in terms of one sequence space versus two,
as discussed by Feldmeier~\cite{feldmeier90multiplexing}
XXX perhaps CM also? any precedent on the ACK overloading issue?}

\xxx{clarify relationship/contrast between DCCP and CM,
	as alternative design points in Flow Layer space:
	- CM allows upcalls at end host to let application
	  make transmit decisions at last possible moment;
	  DCCP provides only the conventional send-queue API.
	- CM is rate-based, doesn't assume or require regular acks;
	  DCCP does acks and is window-based in default CC algorithm.
	- CM can operate to some degree if only one endpoint is CM-aware
	  ...as long as transport provides feedback info?
	- CM doesn't assume every other received packet is acknowledged
	  DCCP probably also supports less-frequent acks in alt mode...?
	- CM congestion control counts both packets and bytes?
	- DCCP supports explicit CC alg negotiation, CM doesn't (yet)
	- CM built around aggregation, CC scoped across all flows
		(like Touch's TCP ctl block interdep RFC 2140)
		(also ensemble TCP)
	they both implement {\em some} arrival/loss-rate feedback mechanism
	independent of anything having to do with reliability.

	tentative suggestion:
	DCCP closer to the protocol we want,
	but would like features of CM:
	e.g., aggregation, and late transmit decision/ALF
}

}
}

\section{Simulation Experiments}
\label{sec-flowsim}

To illustrate how flow splitting can address practical difficulties
caused by network heterogeneity,
we explore two simple but realistic scenarios via simulation.
We implemented a prototype Flow Layer supporting flow splitting
in the ns2 network simulator,
building on existing TCP congestion control algorithms
already supported by the simulator,
and used it to compare relevant performance properties
of flows employing flow splitting against pure end-to-end flows.
These scenarios are intended to illustrate the benefits
of architectural support for flow splitting,
and not to exhaustively analyze or quantitatively predict
real network performance using particular protocols.
We leave analysis of more diverse scenarios and implementation tradeoffs
to future work.

\subsection{Getting Low Delay from Residential DSL}

\begin{figure}[tbp]
\centering
\includegraphics[width=0.47\textwidth]{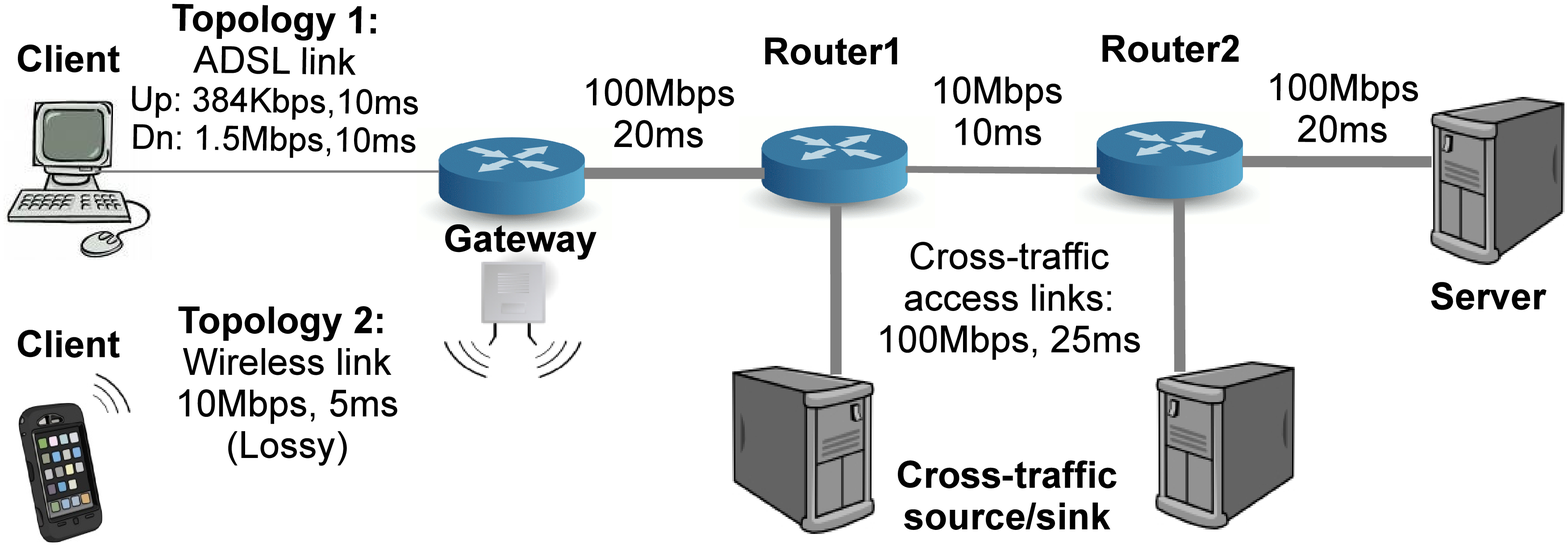}
\caption{Network topology used in simulations}
\label{fig-simtopology}
\end{figure}

We first explore a typical scenario in which a residential DSL connection
is used concurrently for both delay-sensitive activities such as gaming
and bandwidth-intensive activities such as web browsing or file downloads.
The simulation uses the topology shown in Figure~\ref{fig-simtopology} 
(Topology 1),
in which a gateway on the ISP's network
separates the user's client from the Internet.
The client communicates with the server on the far right,
but a pair of hosts generate competing cross-traffic
on an intermediate network link.
We configured the ADSL link
according to observed parameters~\cite{dischinger07characterizing}.

\com{
and we now describe one scenario that we studied. The first topology
in Figure~\ref{fig-simtopology} is motivated by an end-user, such as
an avid gamer or a video-conferencing tele-commuter, who values both
bandwidth and end-to-end delay. ADSL is the access technology, and the
link between Router1 and Router2 is shared with cross-traffic flows
between the cross-traffic nodes. This shared link has a 100ms queue,
the ADSL link has 300ms and 1000ms queues on the downlink and uplink,
respectively (based on observed
queue-sizes~\cite{dischinger07characterizing}), and all other links
have 50ms queues.
}

The ISP in this scenario offers a premium ``gaming service,''
in which the client's gateway acts as a flow middlebox
helping the client maintain low delay.
The client's end host or DSL modem
negotiates the use of a delay-minimizing congestion control scheme
over the DSL link with the flow middlebox---%
we use TCP Vegas~\cite{brakmo95tcp}---%
but the rest of the path from the gateway to the server
uses loss-based NewReno congestion control.
The bottleneck for our observed flow is at the DSL link.

\com{
The Gateway router is owned by the user's ISP and connects the ISP's
network to the wide-area network. In the \tng simulations, the Gateway
is more than a simple router---it is an operator-managed \tng Flow
Middlebox. The operator thus controls the Flow middlebox, which is
setup with a 10-packet queue per flow for uploads, and a 15-packet
queue for downloads.

TCP-NewReno cross-traffic, set up between the
cross-traffic nodes, competes for the Router1-Router2 link
bandwidth. This cross-traffic is in the same direction as the observed
data flow (i.e., from left to right when the observed data flow is an
upload from client to server, and the other way when the observed data
flow is a download from server to client. In all simulations, there is
no cross-traffic at the beginning, and one TCP-NewReno cross-traffic
flow joins in every 250ms. Thus, there are 3 TCP-NewReno cross-traffic
flows alongside our observed flow sharing the Router1-Router2 link
from 750-1000ms.

The two chief contenders currently viable for this scenario are
TCP-NewReno, which is aggressive in getting bandwidth from other
flows, and TCP-Vegas, which is built to maintain low queue-sizes and
thus low end-to-end delay. Since we expect most of the queue build-up
to happen at the deep queues on the ADSL hop, we compare these
contenders with a specific instantiation of the \tng Flow Layer, which
uses a flow middlebox to compose the low-delay TCP-Vegas on the ADSL
link and the aggressive TCP-NewReno in the wide-area.
}

\begin{figure}[tbp]
\centering
\includegraphics[angle=270,width=0.47\textwidth]{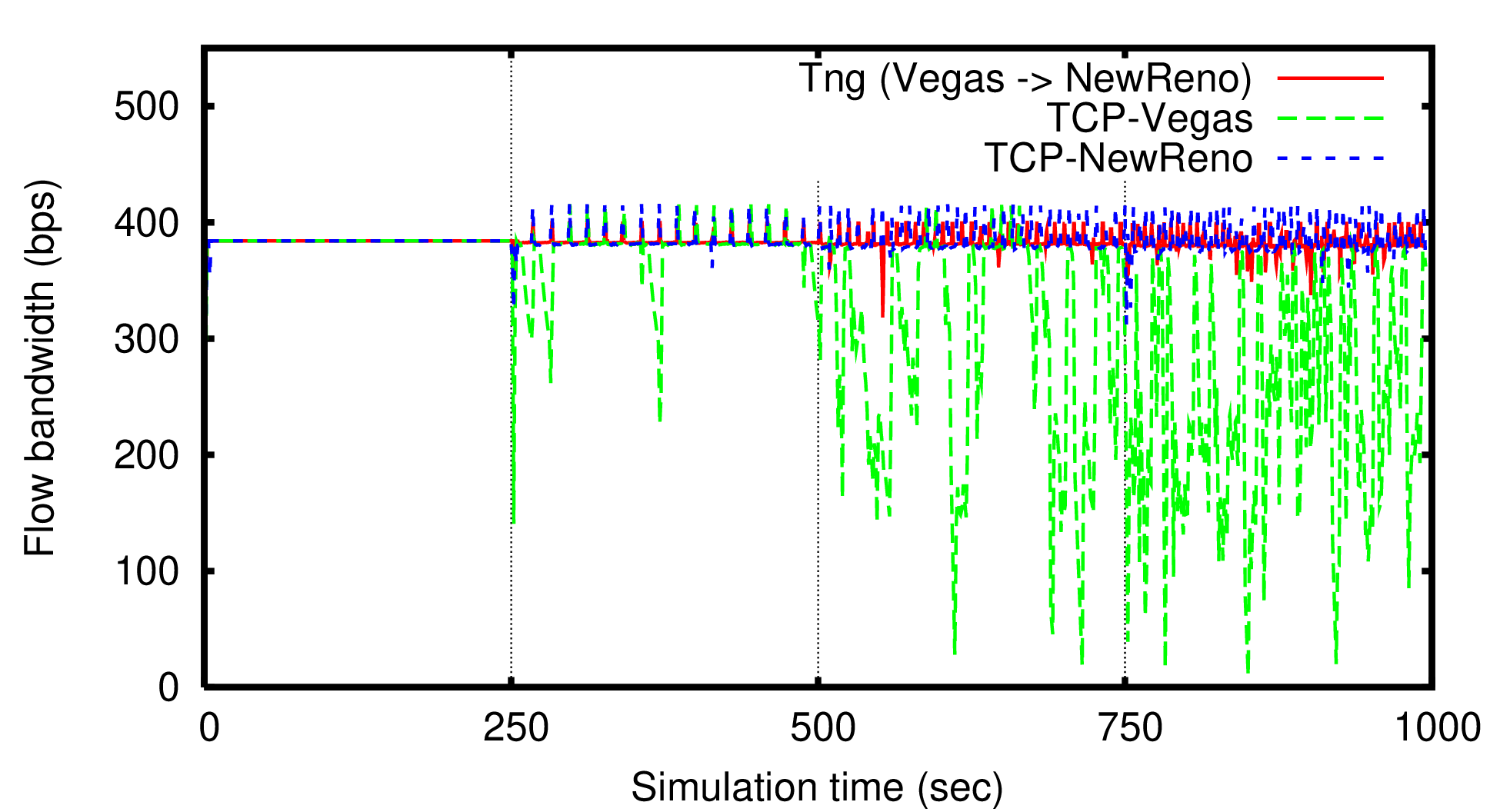}
\includegraphics[angle=270,width=0.47\textwidth]{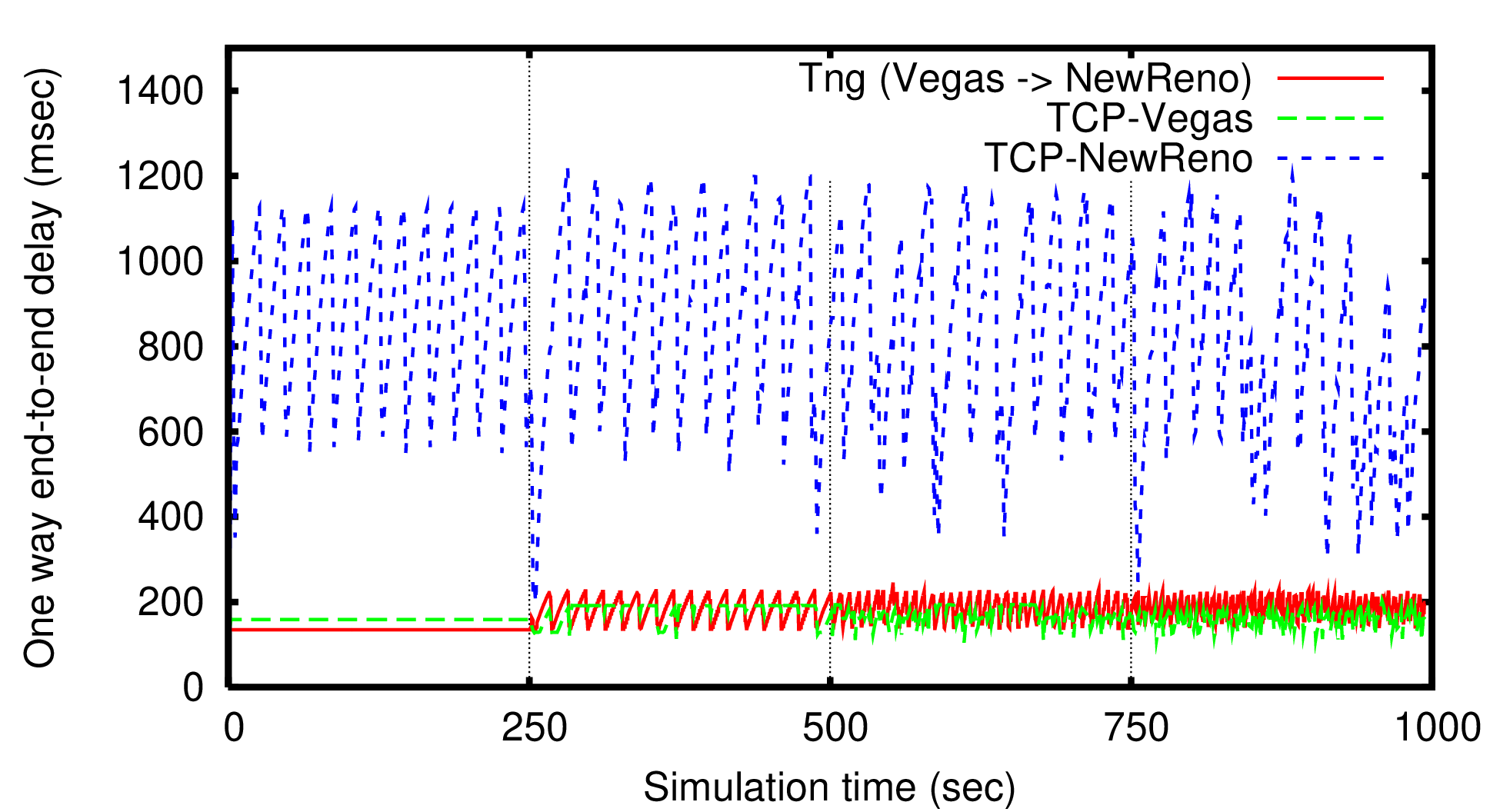}
\caption{(a) Bandwidth obtained and (b) end-to-end delay
  during a DSL upload, measured at 2.5 second intervals over the
  flow's lifetime. One TCP-NewReno cross-traffic flow is added every
  250 seconds.}
\label{fig-dslup}
\end{figure}

Figure~\ref{fig-dslup}
compares the bandwidth and round-trip delay provided
by this \tng-enabled ``gaming service''
against the performance of either NewReno or Vegas alone operating end-to-end,
in the presence of a constant upload stream from the client to the server
and a varying amount of competing cross-traffic on the core Internet.
The simulation adds a new TCP-NewReno cross-traffic flow
every 250 seconds.
As the bandwidth graph shows,
end-to-end Vegas performs well until the first competing NewReno flow appears,
then quickly gives up bandwidth as NewReno cross-traffic increases.
End-to-end NewReno, on the other hand,
competes well with the cross-traffic in securing network bandwidth,
but maintains a consistently high delay---%
a frequent problem for users of
typical DSL modems~\cite{dischinger07characterizing}.
With the \tng-enabled ``gaming service,''
in contrast,
the ISP's flow middlebox isolates the Vegas algorithm
controlling the DSL link
from the NewReno algorithm controlling the path across the Internet core,
enabling the Vegas section to provide low delay
without competing with NewReno flows on the same link,
and enabling NewReno to compete effectively for bandwidth
on the Internet.

\com{
Figure~\ref{fig-dslup} shows bandwidth obtained and delay experienced
by these three solutions for a data upload (client to
server). Figure~\ref{fig-dslup}(a) shows TCP-Newreno saturating the
384 Kbps ADSL uplink quite nicely, but at the cost of saturated queues
at the ADSL uplink, leading to the high end-to-end delays observed in
Figure~\ref{fig-dslup}(b). On the other hand, TCP-Vegas, which is able
to maintain a very small queue-size through the network and is able to
use bandwidth effectively when there's no cross-traffic, competes
poorly for bandwidth as the TCP-NewReno cross-traffic at the shared
link increases. The cross-traffic flows tend to saturate the shared
queue, and TCP-Vegas responds to this queue build-up by unnecessarily
cutting back its sending rate\footnote{In general, delay-based
  congestion control schemes compete similarly in the presence of
  aggressive TCP traffic, and so this analysis applies to other
  delay-based schemes as well.}. The \tng instantiation, which
capitalizes on the strengths of both TCP-Vegas and TCP-NewReno, is
able to both keep end-to-end delays low and compete fairly with the
cross-traffic flows. Figure~\ref{fig-dslup} shows the \tng flow
getting as much bandwidth as TCP-NewReno and experiencing delays as
low as TCP-Vegas.
}

In addition to the main benefit of obtaining low delay while uploading,
the split \tng flow experiences 
slightly lower delay than end-to-end Vegas even without cross-traffic.
This effect results from the shorter feedback loop
that the Vegas client experiences with \tng,
operating over only the ADSL link's 20ms RTT
instead of the full path's 120ms RTT,
an example of the effects
described in Section~\ref{sec-seg}.

\begin{figure}[tbp]
\centering
\includegraphics[angle=270,width=0.47\textwidth]{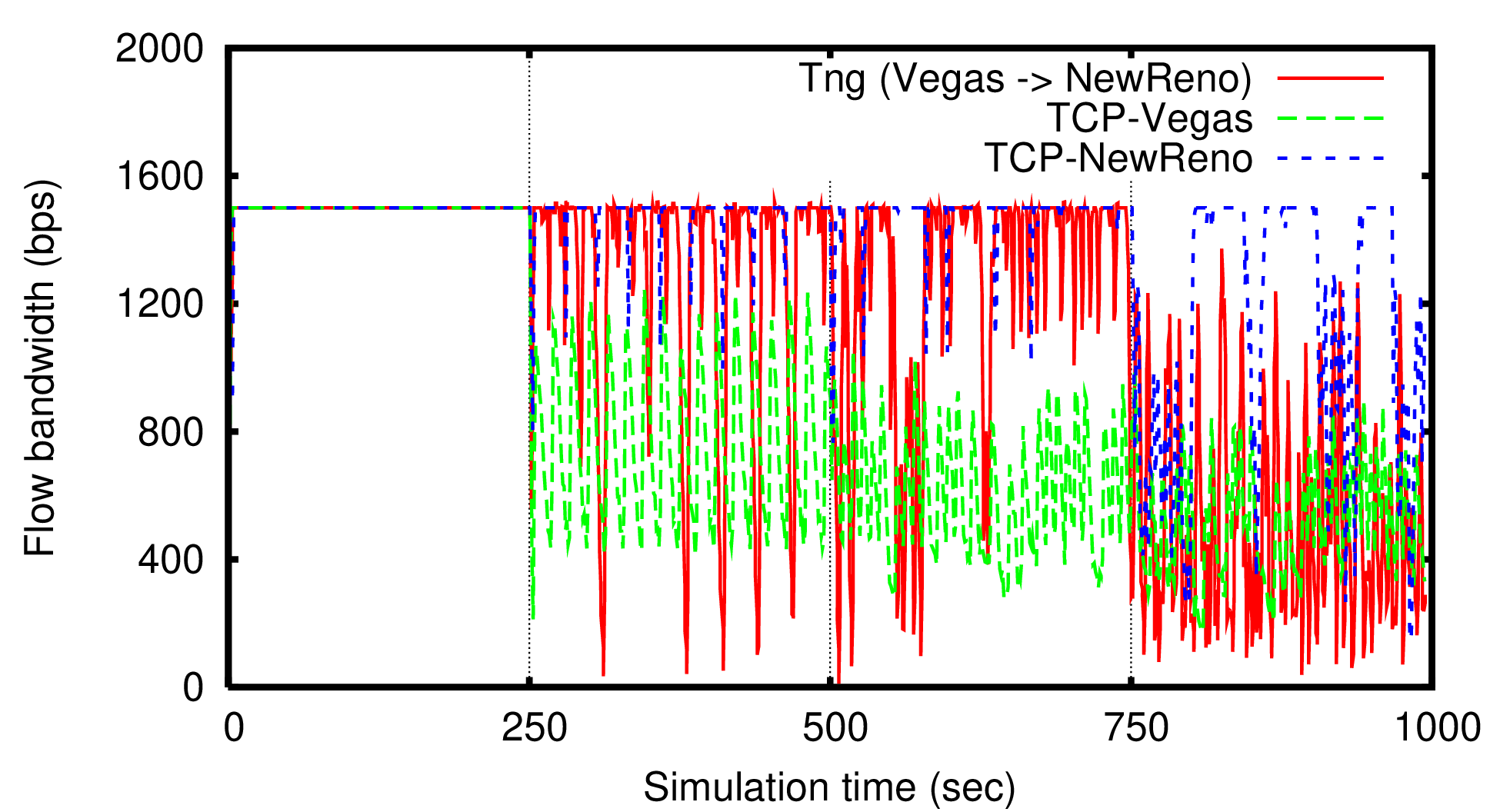}
\includegraphics[angle=270,width=0.47\textwidth]{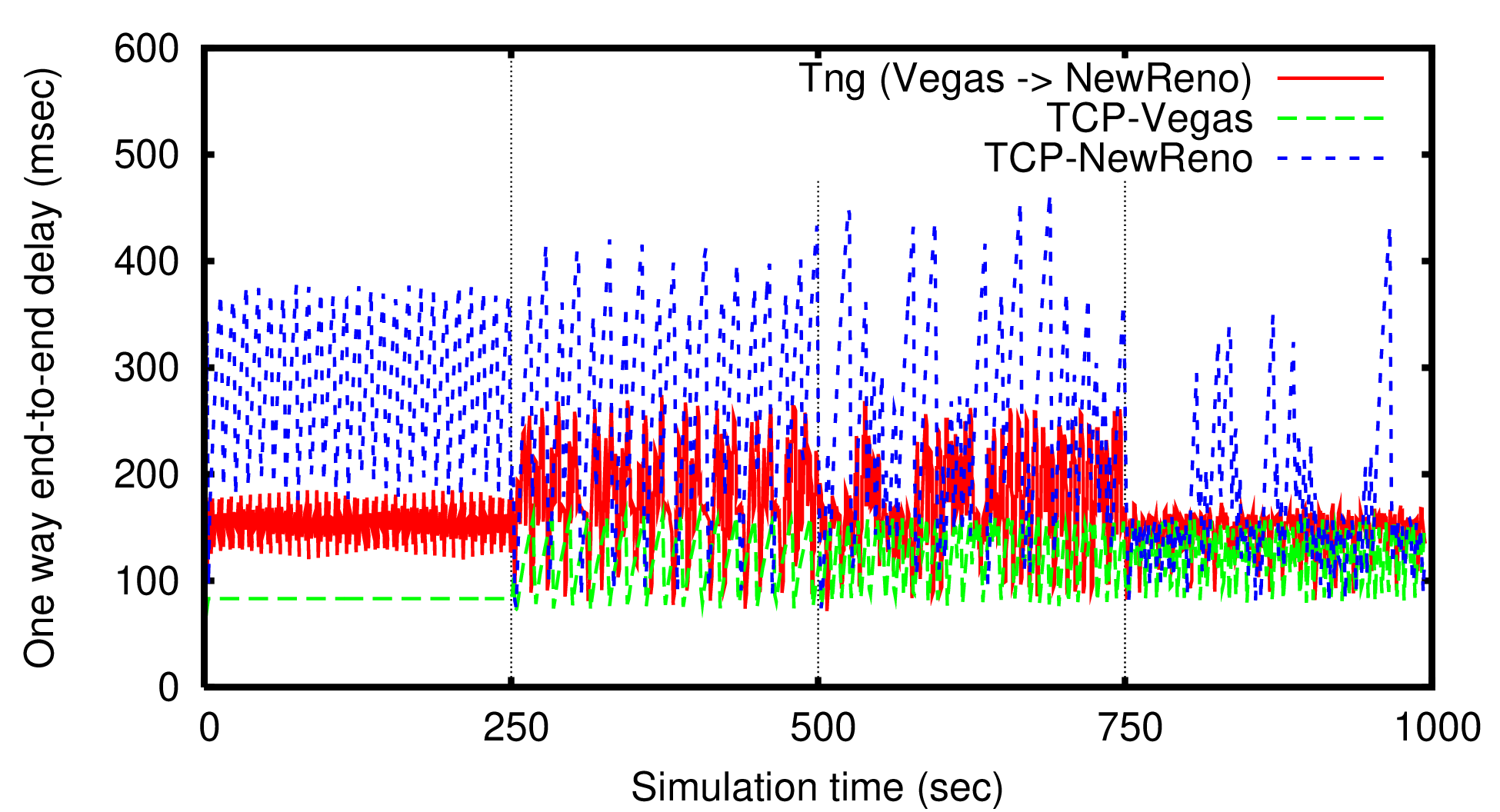}
\caption{(a) Bandwidth obtained and (b) end-to-end delay
  during a DSL download, measured at 2.5 second intervals over the
  flow's lifetime. One TCP-NewReno cross-traffic flow is added every
  250 seconds.}
\label{fig-dsldn}
\end{figure}

Figure~\ref{fig-dsldn}
shows similar results during a download from the server to the client.
The results are similar overall,
but the \tng flow does experience some increase in delay,
though not as much as end-to-end NewReno.
This increase is due to our use of queue sharing
to join Flow Layer sections,
which causes packets crossing from the high-bandwidth NewReno core section
to the lower-bandwidth DSL section
to build up in a NewReno-controlled queue at the flow middlebox
as described in Section~\ref{sec-flow-limits}.
Since this queue is on the high-bandwidth side of the network
and under control of the ISP, however,
it can be made small to serve the low-delay demands of the client.

\com{
Figure~\ref{fig-dsldn}
shows delay and bandwidth dynamics with a data download over
the ADSL topology (server to client). All three solutions are able to
maximize bandwidth when no cross-traffic flows are present (from 0 to
250 seconds), but TCP-NewReno suffers the highest end-to-end delay due
to saturation of the ADSL link and the shared link. TCP-Vegas is able
to maintain a small queue the ADSL link, but witnesses queue build-up
at the shared link as cross-traffic increases, and again, gives up its
bandwidth unnecessarily. \tng shows higher delay than TCP-Vegas but is
not as bad as TCP-NewReno, and it competes as well as TCP-NewReno for
bandwidth in the wide-area.  Cross-traffic increase causes both
TCP-NewReno and \tng to give up some bandwidth, thereby reducing the
queueing delay at the ADSL link, and an already struggling TCP-Vegas
deteriorates further.

There is a tussle here: delay on the downlink is in conflict with
bandwidth. This tussle is centered at the flow middlebox queue, which
was mentioned earlier in Section~\ref{sec-flow-def}: a larger flow middlebox
queue allows TCP-NewReno to better fill the wide-area pipe, resulting
in higher bandwidth for the flow. At the same time, this larger buffer
increases the delay for the end-to-end \tng flow. Viewed differently,
this tussle is a tradeoff that an operator can leverage: A customer
can be offered a ``low delay'' service which can be provided through
control of this queue size knob.
}

Overall, this instantiation of \tng combines the strengths of the
different TCP variants in their specific domains, and thus provides a
high-bandwidth, low-delay service that none of the end-to-end
schemes could manage alone.

\subsection{A Lossy Wireless Network}

\begin{figure}[tbp]
\centering
\includegraphics[angle=270,width=0.47\textwidth]{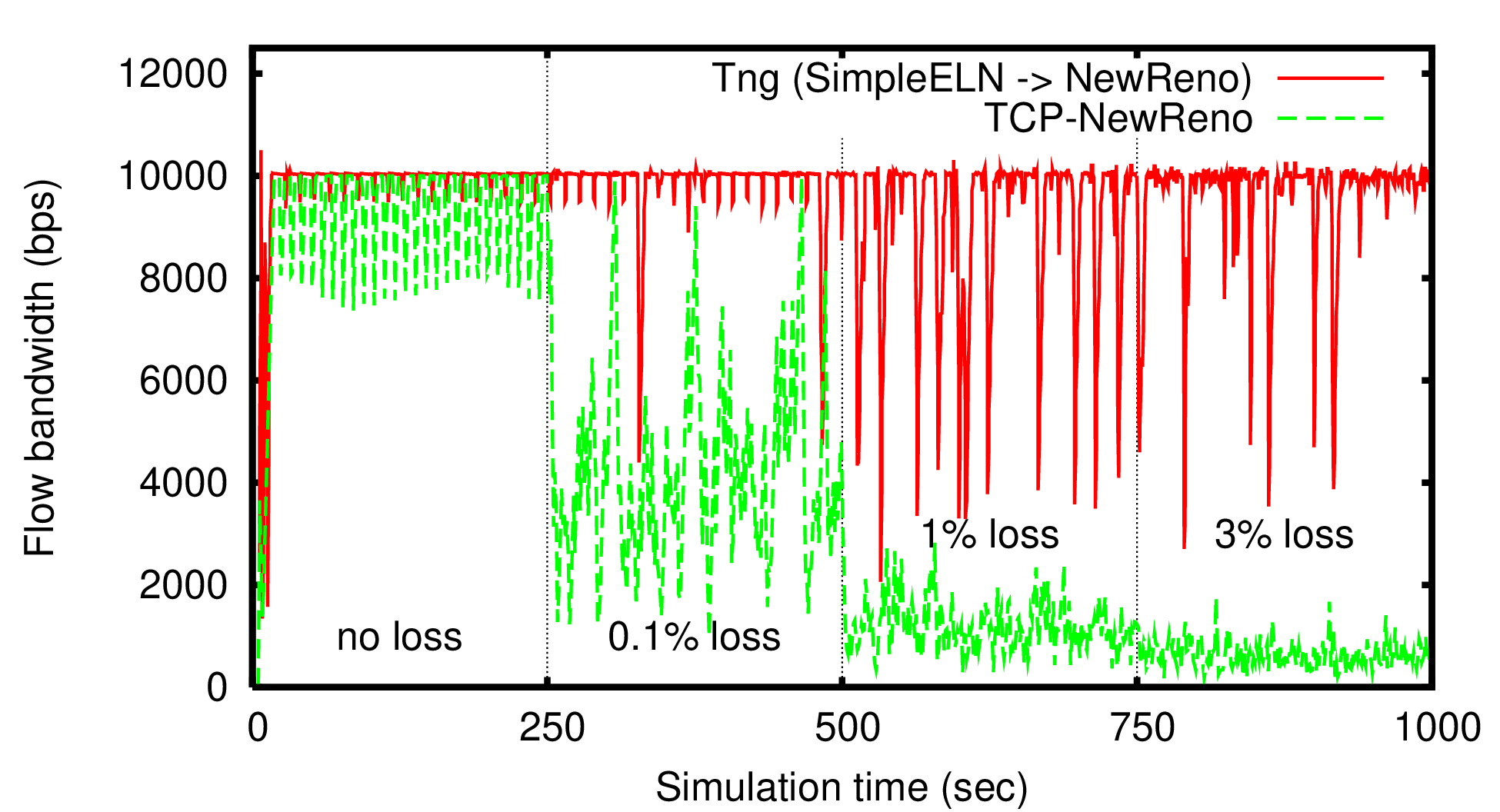}
\includegraphics[angle=270,width=0.47\textwidth]{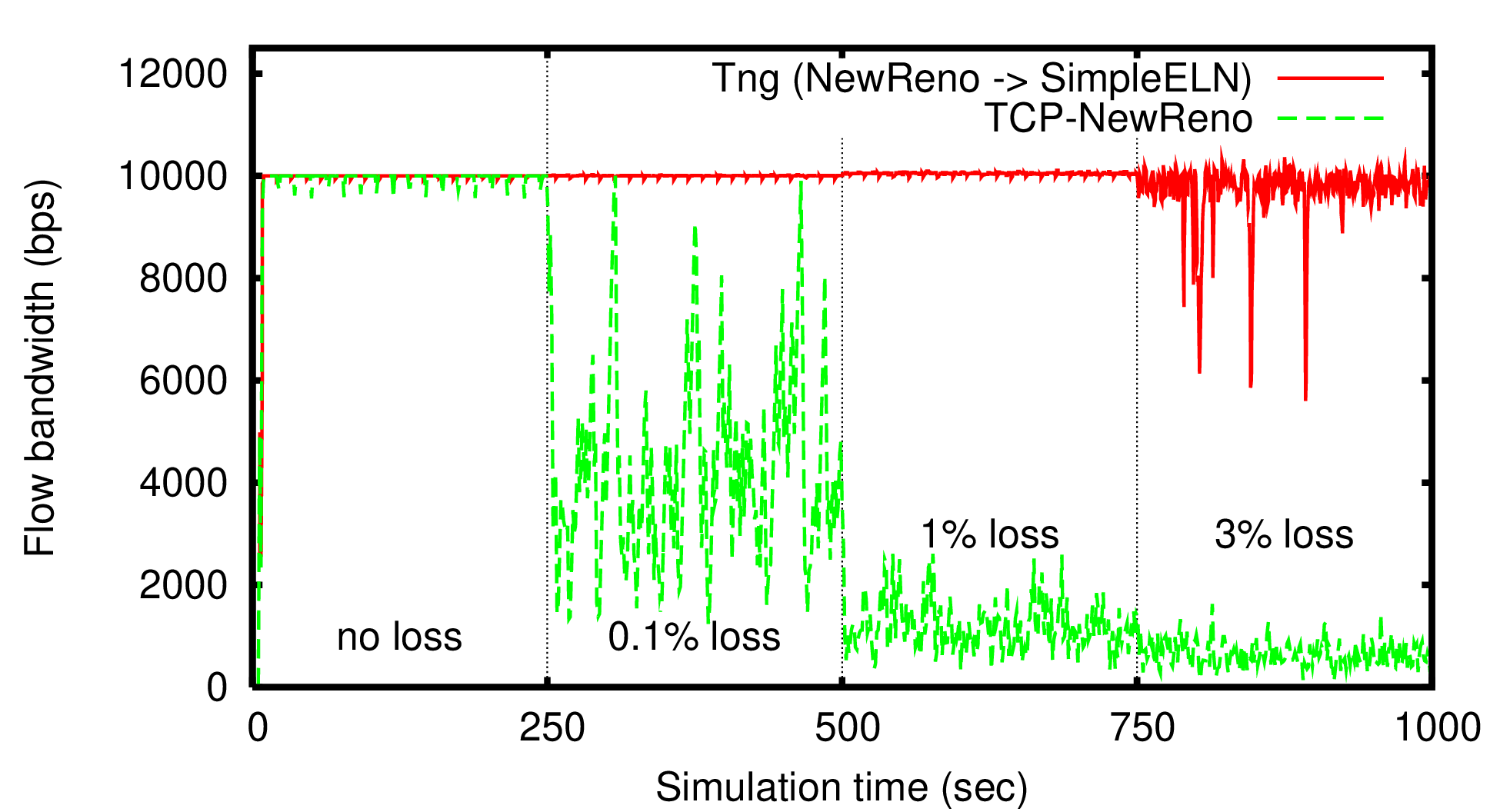}
\caption{Bandwidth obtained by data (a) upload and (b) download
  flows over the lossy wireless topology, measured over 2.5 second
  intervals, over the flow's lifetime.}
\label{fig-eln}
\end{figure}

The second topology in Figure~\ref{fig-simtopology} uses a wireless
link at the last hop with a varying loss rate. This topology is
motivated by a mobile/wireless end-user who is chiefly concerned with
maximizing bandwidth.

We implemented TCP-SimpleELN, a TCP variant supporting
Explicit Loss Notification (ELN)~\cite{balakrishnan98explicit}
signals from the TCP-SimpleELN
receiver. The TCP-SimpleELN receiver accepts notifications of
packet loss from the underlying wireless link layer. When such a
notification is received, the TCP-SimpleELN receiver sends back a
message to the sender explicitly indicating packet(s)
that were dropped by the link layer. The TCP-SimpleELN sender then
retransmits the dropped packet(s) without modifying the
congestion window.

Figure~\ref{fig-eln} shows the performance of end-to-end TCP-NewReno
and an instantiation of \tng composed of TCP-SimpleELN on the last
wireless hop and TCP-NewReno in the wide-area. The loss rate increases
from 0 at the beginning to 0.1\% at 250 seconds, then to 1\% at 500
seconds, and finally to 3\% at 750 seconds. \tng is able to leverage
TCP-SimpleELN's strength on the wireless link, and maximizes bandwidth
for both data uploads and downloads.

Since TCP-SimpleELN relies on a link layer notification, the transport
receiver must be co-located with the wireless link layer
receiver. \tng makes this possible for any end-to-end flow, since the
lossy link layer can be managed by flow middleboxes using
TCP-SimpleELN on the link. 



\section{A Prototype \TNG Stack}
\label{sec-impl}

While Section~\ref{sec-flowsim}'s simulations
suggest the feasibility of joining flow sections via queue sharing,
we wish to evaluate flow splitting
in the context of the overall \tng architecture
to validate our original goal
of supporting in-path optimization
without interfering with end-to-end transport functions.
To do so,
we built a prototype protocol suite
demonstrating the proposed refactoring of transport services
into Endpoint, Flow Regulation, Isolation, and Semantic Layers,
thereby achieving \tng's main goals.
This section describes relevant details of our current prototype
together with experiments using the prototype 
that confirm \tng's feasibility and illustrate the benefits
of its clean support for flow splitting.

\com{
Since this paper takes a long-term perspective
on refactoring the Internet architecture
implementing a complete and production-ready protocol suite
is far beyond the scope of this paper.
We nevertheless wish to validate the crucial elements of the architecture
and test their technical viability.
To do so,
we built a prototype protocol suite that,
while implementing only a subset of the architecture we envision,
demonstrates the proposed re-factoring of transport services
and achieves \tng's main goals.
This section describes relevant details of our current prototype
together with experiments using our prototype 
that confirm the feasibility and illustrate the usefulness
of \tng's critical components.
}

\com{
One of the short-term benefits of the proposed architecture
is that existing protocols already provide starting points
for implementing its new layers.
Since these existing protocols were designed
in the traditional architectural framework, however,
the fit is not perfect,
so further development will be needed.

Although the layering model proposed in this paper
focuses on functionality (XXX clarify)
for separability and composability purposes,
this functional layering does not necessarily need to imply
that each layer is implemented completely independently.
In fact, in our prototype implementation,
we apply the principle of XXX (Joe Touch's RNA stuff)
by implementing both the Flow Layer and the Isolation Layer
through different instantiations of a single protocol.
This approach has both advantages and disadvantages...
}

\subsection{Organization of the Prototype}
\label{sec-proto-org}

\begin{figure}[t]
\centering
\includegraphics[width=0.45\textwidth]{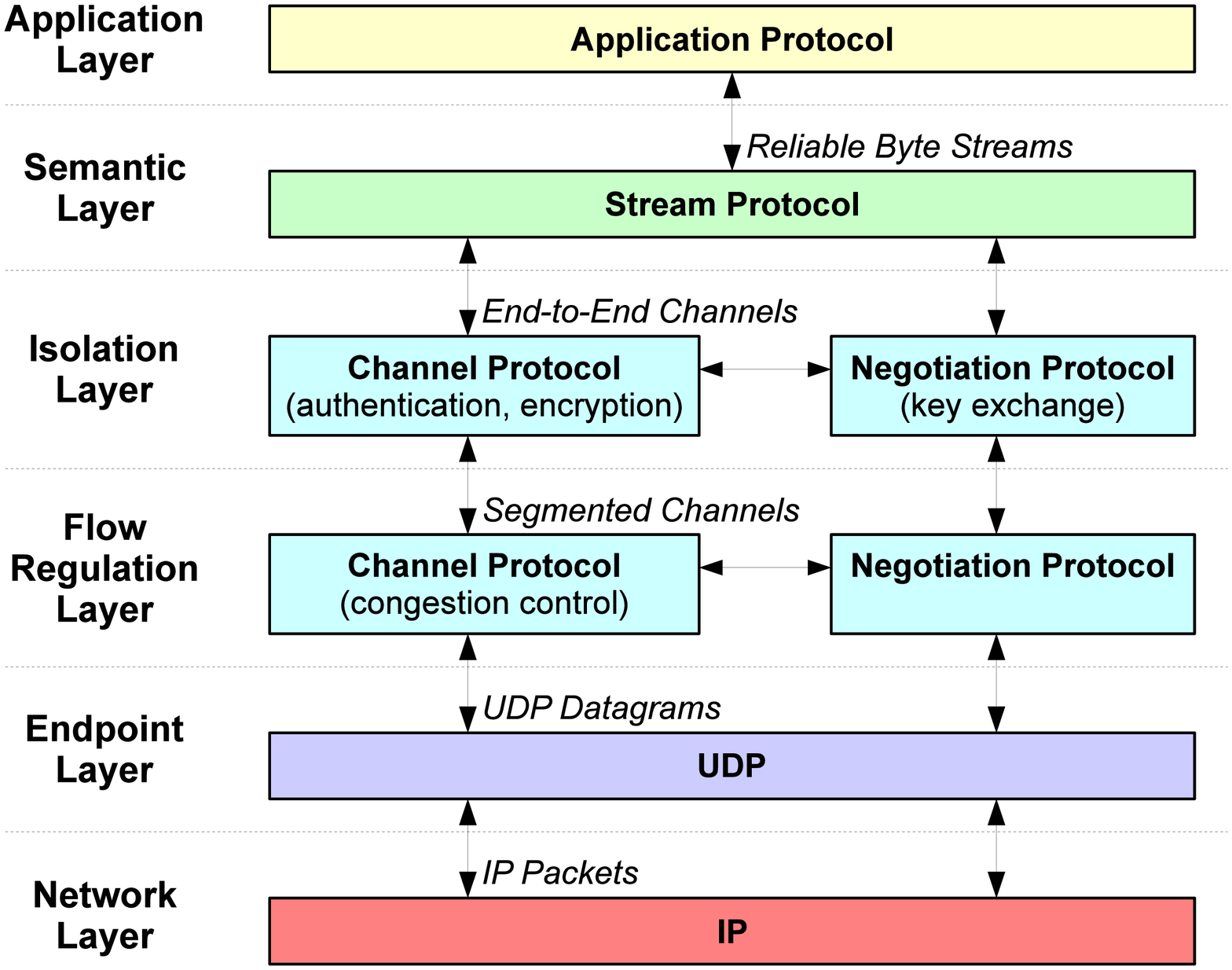}
\caption{Protocol Design of the Prototype}
\label{fig-prototype}
\end{figure}

Figure~\ref{fig-prototype} illustrates
the overall structure of the prototype,
which builds on a previous experimental prototype
of the Structured Stream Transport (SST) protocol~\cite{ford07structured}.
SST consists of two main components:
a Channel Protocol and a Stream Protocol.
The Channel Protocol implements a sequenced and congestion-controlled
but unreliable and unordered packet delivery service,
comparable to DCCP~\abcite{rfc4340}{kohler06dccp},
but with optional cryptographic authentication and encryption
similar to that of IPsec~\cite{rfc4301} and DTLS~\cite{rfc4347}.
The Stream Protocol builds on
the Channel Protocol's delivery service
to provide reliable, ordered byte streams semantically equivalent to TCP's,
but capable of being created and destroyed more efficiently,
enabling fine-grained (e.g., transactional) use of these lightweight streams.
This separation of functions within SST is the reason for it being
the basis of our prototype:
SST's Stream Protocol nicely fits the role of \tng's Semantic Layer,
its Channel Protocol, while needed to be reworked as described below, serves as 
starting point for both \tng's Flow and Isolation Layers,
and its Channel Protocol already builds atop UDP
as a starting point for \tng's Endpoint Layer.

The main challenge was implementing
the Flow Regulation and Isolation Layers.
To do so, we borrowed a principle of
the Recursive Network Architecture~\cite{touch06recursive},
and adapted the Channel Protocol so that this one protocol
may be instantiated in different configurations
to implement both the Flow Layer and the Isolation Layer.
When implementing the Flow Layer,
the Channel Protocol operates with congestion control enabled
but cryptographic security disabled,
and we modified the protocol to allow
dividing an end-to-end path into segments,
each running a separate instance of the Channel Protocol
with an independent congestion control loop.
When implementing the Isolation Layer,
the Channel Protocol operates end-to-end,
using self-certifying cryptographic identifiers
as in HIP~\cite{rfc4423}\abbr{}{ and UIA~\cite{ford06persistent}}
to give hosts stable identities
as they migrate among IP addresses,
and using IPsec-like encryption and authentication
to secure the end-to-end channel against interposition or eavesdropping.
The end-to-end channel serving as the Isolation Layer
runs with its own congestion control logic disabled,
relying instead on the underlying, segmented Flow Layer
instance(s) of the Channel Protocol to implement this function.

\xxx{ installed base...
is also deployable by applications
without requiring special privileges or kernel extensions,
and is compatible with 
in the presence of 
The prototype also demonstrates that \tng can be implemented
in a way that is fully backward-compatible with 
}

\com{
The Channel Protocol's
design follows that of IPsec in many respects, providing sequencing
and replay protection along with packet authentication and encryption.
Unlike IPsec, the Channel Protocol also includes selective packet
acknowledgment similar to that of DCCP; while Flow Layer instances of
the Channel Protocol use this feedback service to implement congestion
control, an Isolation Layer instance of the Channel Protocol performs
no congestion control but still includes the feedback function as a
service to the Stream Protocol.
}

The Stream Protocol
does not require a stream to be attached always to the same channel:
instead, a stream can {\em attach} dynamically
to any available channel between the appropriate pair of hosts,
as identified cryptographically by the Isolation Layer.
Each Flow Layer channel monitors the channel's condition
using the same packet-level acknowledgments
it uses to implement congestion control,
and reports its condition to higher layers.
\com{
whenever the flow's state changes
between ``up,'' ``uncertain,'' and ``down.''
A flow's state becomes ``up'' when a new transmission is acknowledged,
``uncertain'' when a retransmission timeout occurs,
and ``down'' after too many unacknowledged retransmissions.
}
If a flow detects a stall or failure,
the Isolation Layer channel atop that flow
propagates this signal upward to the Semantic Layer,
which attempts to construct Flow and Isolation Layer channels
representing a new or alternative communication path.
If a new, authenticated end-to-end channel comes online
while the old one is still unusable,
the Stream Protocol migrates existing streams to the new channel
transparently to the application.

Associated with the Channel Protocol,
SST uses a separate {\em Negotiation Protocol} for key exchange,
similar to IPsec's IKE~\cite{rfc4306}
or HIP's key exchange mechanism~\cite{rfc5201}
and based on Just Fast Keying~\cite{aiello04jfk}.
Finally, to enable hosts to find each other after changing IP addresses,
SST provides a simple {\em Registration Protocol}
analogous to a name service
through which hosts can register their cryptographic identities
with a registration server
and look up the current network endpoints of other hosts
by their cryptographic identities.

The prototype protocol suite runs in user space,
and is implemented in C++ using the Qt event framework~\cite{trolltech-qt}.
It includes an asynchronous networking framework
that enables it, and applications using it,
to be run either on real networks
or in a network simulation environment
for development and testing purposes.
When used in the simulation environment,
the protocol suite still implements complete, working protocols
that exchange and process ``real'' packets containing user data,
so it is more faithful in this respect than many simulation environments.
\com{	XXX do this\ldots
}

\com{
The following sections describe relevant protocol design issues
for each of the four architectural layers in more detail,
and how the current prototype addresses those issues.
}

\com{Our Isolation Layer implementation  uses an end-to-end
instance of SST's Channel Protocol configured for strong security,
which combines the functions of IPsec and HIP by providing both
cryptographic communication channel protection based on
self-certifying cryptographic host identities.  
}

\subsection{Validating Flow Splitting in the Prototype}
\label{sec-impl-split}

\begin{figure}[t]
\centering
\includegraphics[width=0.48\textwidth]{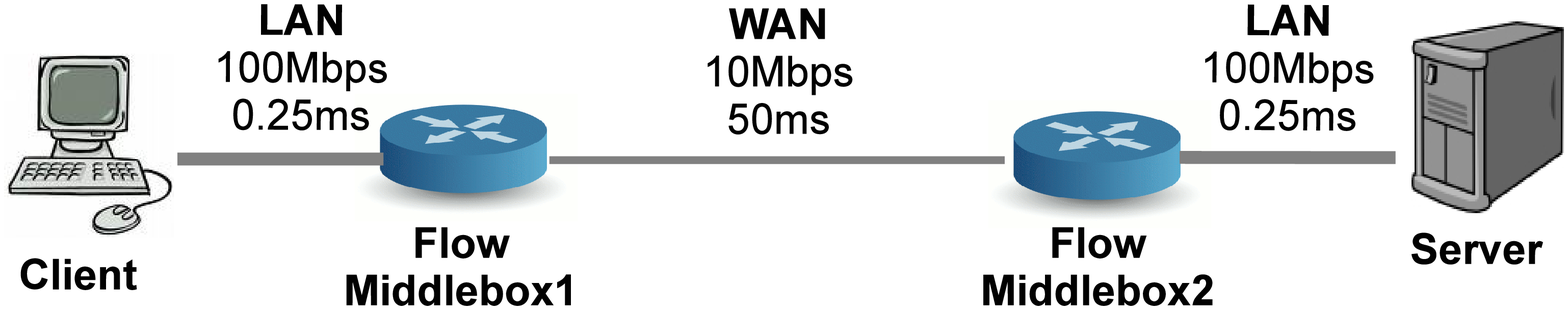}
\caption{Experimental topology for long-delay inter-site link scenario.}
\label{fig-intersite}
\end{figure}

\begin{figure}[t]
\centering
\includegraphics[width=0.48\textwidth]{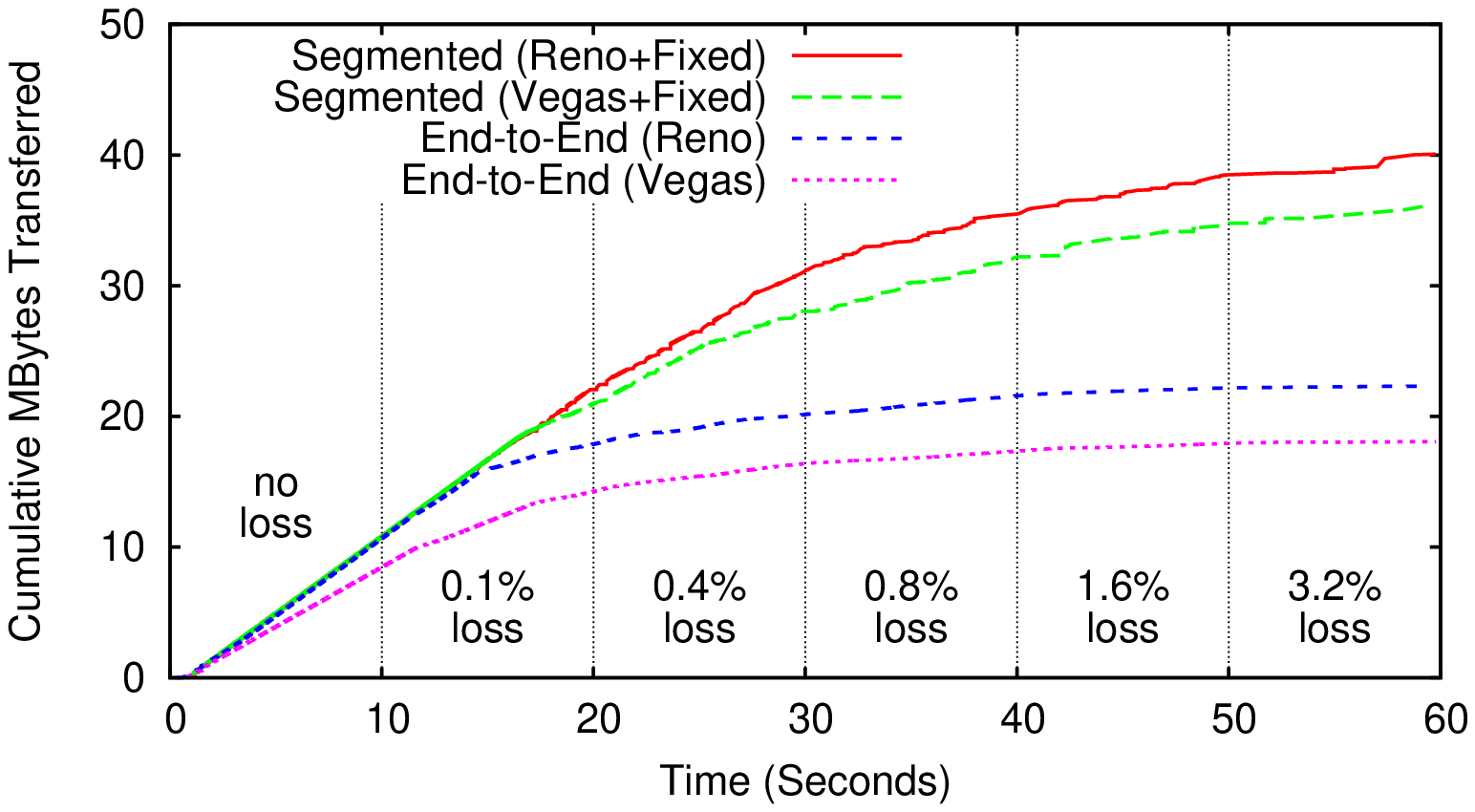}
\caption{End-to-End reliable transfer performance
	over a high-bandwidth-delay-product link with random loss,
	with and without flow splitting.}
\label{fig-protobytes}
\end{figure}

To validate flow splitting via the prototype's Channel Protocol,
we test a simple network scenario
corresponding to a common use of PEPs
around a high-bandwidth, long-distance link
such as a reserved-bandwidth link between two sites
in an organization's private network.
To simplify experimentation and provide exactly reproducible results,
we run the protocol suite in the prototype's network simulation environment.
The experiment uses
the simulated network topology shown in Figure~\ref{fig-intersite},
consisting of two high-bandwidth, low-delay LAN links
surrounding a medium-bandwidth, high-delay WAN link,
with the WAN link incurring a variable random loss rate.

In the \tng version of the scenario,
the flow middleboxes surrounding the link
interpose on Flow Layer sessions traversing the link
to optimize flow performance.
Since this inter-site link provides fixed point-to-point bandwidth,
we assume that the WAN link itself
needs no congestion control---%
only the LANs on both ends do.
The WAN section runs a trivial ``congestion control'' scheme
that merely maintains an administratively fixed transmission rate
corresponding to the link's bandwidth.
This way a flow using the section takes no time to ramp up
to full use of the section,
and there is no need for special techniques to
distinguish congestion from non-congestion losses
since there are no congestion losses.
Of course, to share the link among multiple flows
the middlebox must divide the link's fixed congestion window
among the flows,
similar to XCP's fairness controller~\cite{katabi02internet}.

Figure~\ref{fig-protobytes} plots cumulated bytes transferred over time
by a long reliable data transfer using the Stream Layer,
over the \tng-split flow versus an equivalent end-to-end flow,
using both Reno-like and Vegas-like congestion schemes.
We plot cumulative bytes in this experiment instead of average bandwidth
because the Stream Protocol's byte stream reordering
creates violent artificial spikes in a bandwidth plot.
Every 10 seconds in the simulation,
the WAN link's random loss rate increases.
This loss quickly affects end-to-end throughput
as both Reno and Vegas misinterpret the random loss as congestion loss,
but in the split scenario the flow middleboxes
shield the endpoints and the LAN sections from these loss effects,
resulting in good performance until the loss rate becomes very large.

\xxx{explain why we don't use bandwidth graph}

\xxx{drive point home: what's interesting is not that this works;
it's the same thing that many other indirection protocols have done.
what's interesting is that it does it underneath and independently 
of the isolation and semantic layer.}

\subsection{Recovering from Flow Layer Failures}
\label{sec-impl-fate}

While conventional PEPs
might implement the optimizations
described in the previous experiments,
\tng's key novelty
is its support for such optimizations
without their interfering with end-to-end security or reliability.
Section~\ref{sec-impl-split} already offers ``proof by example''
of flow splitting coexisting with end-to-end security,
as the Isolation Layer channel provides end-to-end security
while running atop
multiple per-section Flow Layer channel instances.

\begin{figure}[t]
\centering
\includegraphics[width=0.48\textwidth]{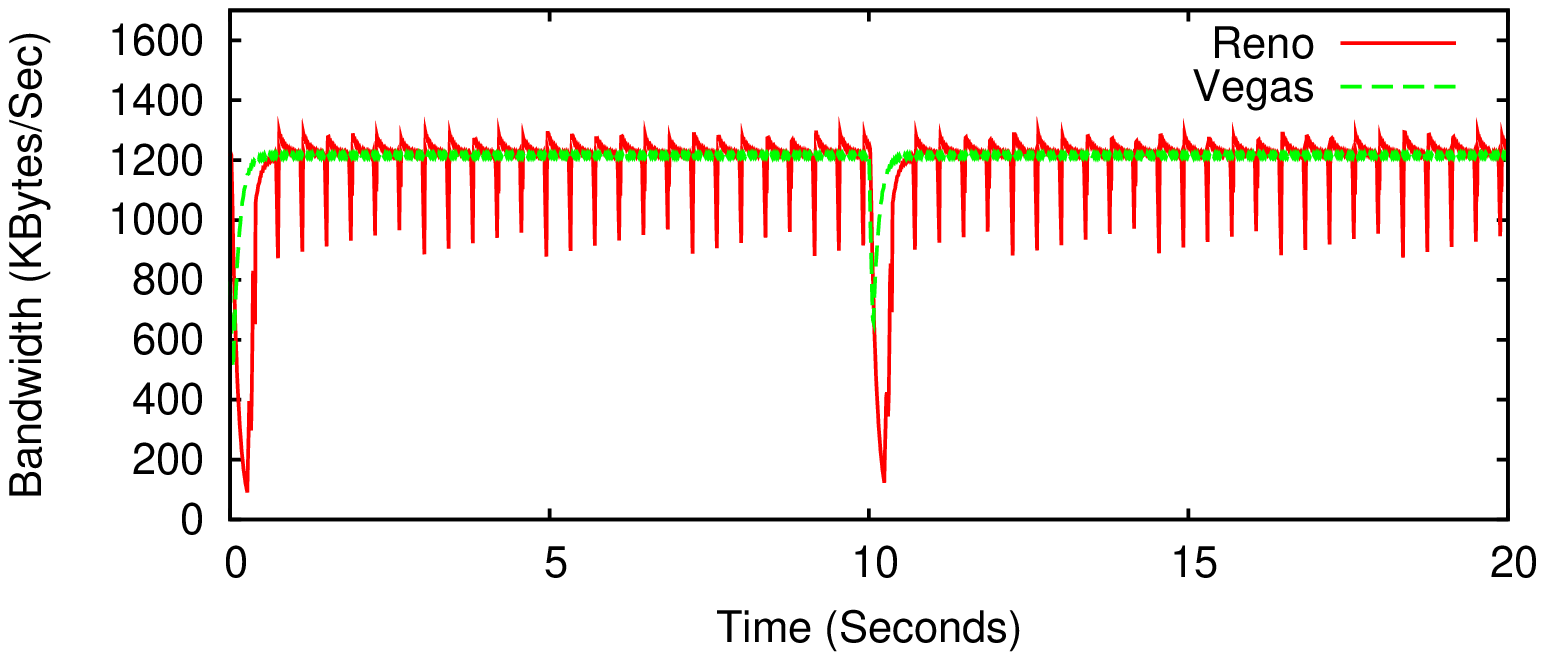}
\caption{Bandwidth trace of an end-to-end data transfer
	across a migration event using the \tng prototype:
	the sending host changes its IP address at 10 seconds.}
\label{fig-migr}
\end{figure}

To demonstrate \tng's preservation
of end-to-end reliability~\cite{saltzer84endtoend}
and fate-sharing~\cite{clark88design}
despite Flow Layer failures or network reconfigurations,
as argued in Section~\ref{sec-flow-def},
we now test the prototype in a simple migration scenario.
Figure~\ref{fig-migr} shows a trace
of an end-to-end, application-level data transfer
using the prototype
over a simulated 10Mbps link,
where the IP address of one of the endpoints (the sender in this case)
changes 10 seconds into the trace.
Once the Flow Layer's congestion control loop
detects and reports a stall
as described in Section~\ref{sec-proto-org},
the Semantic Layer initiates the construction
a new set of Flow and Isolation Layer channels to the remote host,
which includes a new Registration Protocol query
to find the host's latest IP address.
As the figure indicates,
the prototype requires only a few round-trips after the stall
to find the host's new IP address
and negotiate new end-to-end encrypted and authenticated channels,
before migrating and resuming the stream transparently to the application.

If the link or network layer could provide
advance warning of an impending network reconfiguration,
and permit simultaneous use of the new and old network configurations
during a transition period,
then \tng could mask even this temporary interruption
by negotiating new channels while continuing to use the old ones.

\com{
For the Semantic Layer, we use the SST Stream Protocol,
which is designed to run atop the Channel Protocol:
e.g., it takes advantage of the sequencing
and end-to-end acknowledgment
the Channel Protocol provides
in order to enable fast connection setup and teardown.
This aspect of the protocol design
is not essential to our layering model, however;
we expect that other more traditional transports
such as TCP or SCTP could easily be ported to our architecture
merely by disabling their congestion control functionality
when they are running atop a Flow Layer protocol.
}

\section{Deployment Strategies}
\label{sec-deploy}
\com{
This paper explores flow splitting and Flow Middleboxes
in the larger context of \tng;
we now examine some alternative strategies for deployment
of \tng, and thus flow splitting.
}
Any refactoring of existing Internet protocols
faces major deployment hurdles due to the Internet's inertia,
and \tng is no exception.
However,
we find several reasons for optimism
that an architecture incorporating the principles described here
could overcome these deployment hurdles.
Specific strategies that can facilitate \tng's deployment follow.

\com{
\subsection{Protocol Suite Design Alternatives}

While our prototype uses new protocols
for all layers except the Endpoint Layer,
it should also be possible to implement our architecture
by adapting more established protocols.
DCCP~\abcite{rfc4340}{kohler06dccp} or
CM~\cite{balakrishnan99integrated},
modified to support flow middleboxes and flow splitting,
could implement our Flow Layer;
IPsec~\cite{rfc4301} and/or HIP~\cite{rfc4423},
modified to run atop the Flow Layer,
could provide our Isolation Layer;
and TCP or another existing transport,
modified to disable its internal congestion control functions
and instead use those of the underlying Flow Layer,
could provide our Semantic Layer.

There would be costs to this approach:
the existing protocols were not designed for \tng,
and they each independently re-implement considerable functionality
that our prototype reuses or shares synergistically between layers.
Since we instantiate one Channel Protocol in different configurations
to implement the Flow and Isolation Layers, for example,
most of the protocol's code is shared between the two layers,
such as sequencing, replay protection, and acknowledgments.
Because SST's Stream Protocol is designed to run only atop a Channel Protocol
that provides sequencing and end-to-end acknowledgment,
the Stream Protocol is more lightweight
in both code and per-packet overhead
than it would be if it had to reimplement these functions.
}

\begin{table}[t]
\centering
\begin{small}
\begin{tabular}{l||l|l||r|r||r|r}
		& \multicolumn{2}{|c||}{Protocols}
			& \multicolumn{2}{|c||}{Header Size}
				& \multicolumn{2}{|c}{Code Size}
					\\

Layer		&SST	&Legacy	&SST	&Legacy	&SST	&Legacy	\\
\hline
\hline
Semantic	&Stream	&TCP	&8	&20 	&1600	&5300	\\
\hline
Isolation	&Channel &ESP	&24	&32	&\multirow{2}{*}{930} &5300	\\
\cline{1-5}\cline{7-7}
Flow		&Channel &DCCP	&12	&16	&	&2900	\\
\hline
Endpoint	&UDP	&UDP	&8	&8	&600	&600	\\
\hline
\hline
Total		&	&	&52	&76	&3130	&14100	\\
\end{tabular}
\end{small}
\caption{Protocols, per-packet header overhead,
	and approximate code size (semicolons)
	of SST-based prototype versus comparable legacy protocols
	from Linux-2.6.28.2.
	IPsec/ESP and SST use AES-CTR encryption~\cite{rfc3686}
	with HMAC-SHA256-128 authentication~\cite{rfc4868}.}
\label{tbl-implalt}
\end{table}

\com{
Table~\ref{tbl-implalt} illustrates these tradeoffs
between reuse and efficiency,
by showing the protocols, minimal header sizes, and approximate code sizes
for each implementation alternative.
The ``legacy'' alternative is based on
TCP with no header options for the Semantic Layer,
IPsec's Encapsulating Security Payload (ESP) for the Isolation Layer,
DCCP for the Flow Layer,
and UDP for the Endpoint Layer.
We assume
AES-CTR encryption with HMAC-SHA256-128 authentication
for the Isolation Layer in both SST and ESP\abbr{}{~\cite{rfc4868,rfc3686}}.
Across all four layers,
the SST-based protocol suite saves 24 bytes per packet
in comparison with the legacy design.

Table~\ref{tbl-implalt} shows approximate code sizes in semicolons
for the SST-based prototype,
and for the code in Linux-2.6.28.2
implementing the corresponding legacy protocols.
The code counted does not include Linux's TCP congestion control
(since only DCCP's congestion control would be used in \tng),
key exchange, or cryptographic primitives.
We include these figures for illustrative purposes only:
the prototype is new, experimental, and written mostly in C++,
while the existing protocol implementations are mature and written in C.
}

\com{
A mature kernel implementation of our protocol suite in C
would probably also be much larger than the current prototype.
(Bringing our protocol suite to production readiness
might also require increasing the header sizes of some of the layers,
though we hope not by much.)
}

\com{	Calculations:
	ESP header size:
		8	ESP packet fixed header (RFC4303)
		8	IV for AES counter mode (RFC3686)
		16	Authenticator for HMAC-SHA256-128 (RFC4868)

SST:
	Stream Protocol:
		1645	stream.cc stream.h strm/*.cc strm/*.h
	Channel Protocol
		925	chk32.cc chk32.h flow.cc flow.h sock.cc sock.h 
			timer.cc timer.h util.cc util.h
		which includes:
			122	AESArmor stuff in flow.*
			57	congestion control switch statements;
				but a lot of congestion control functionality
				is shared and scattered throughout.

Linux-2.6.28.2 kernel:
	TCP code size:
		4900	tcp.c tcp_cong.c tcp_input.c tcp_ipv4.c 
			tcp_minisocks.c tcp_output.c tcp_timer.c
		406	include/net: tcp.h tcp_states.h
	ESP code size:
		583	include/net: esp.h xfrm.h
		29	include/crypto: aead.h authenc.h
		293	net/ipv4/esp4.c
		231	crypto/aead.c
		237	crypto/authenc.c
		3932	xfrm/*.*
}


{\bf Existing Protocol Reuse:}
A protocol stack supporting clean flow splitting as in \tng
could be composed entirely of existing protocols:
TCP with congestion control disabled as the Semantic Layer,
IPsec as the Isolation Layer,
DCCP as the Flow Layer, and
UDP as the Endpoint Layer.
This approach may not yield the most far-reaching benefits,
and may incur overheads due to redundancies between layers:
e.g., Table~\ref{tbl-implalt}
compares the minimal per-packet overhead of this reuse approach
against our \tng prototype for comparable functionality,
as well as approximate source code line counts.
Nevertheless, reuse could mitigate the difficulty
of new protocol development and standardization.

{\bf Application Transparency:}
Our \tng prototype's Semantic Layer 
already provides
a reliable stream abstraction compatible with TCP's:
with careful design,
a kernel implementation of \tng
could replace TCP completely transparently to applications,
dynamically probing the network and remote host for \tng support
and falling back on TCP if necessary.

{\bf Compatibility with Existing PEPs:}
While a DCCP-like protocol is most suited to \tng's Flow Layer,
a \tng stack might support
the use of standard TCP as a fallback ``Flow Layer,''
atop which the \tng stack's {\em true}
Isolation and Semantic Layer protocols would run
as if a TCP ``application.''
While TCP's overhead and ordering constraints may incur a performance cost,
encapsulation in legacy TCP flows would make the new stack
even more compatible with existing networks
and capable of benefiting from existing TCP-based PEPs,
and could still restore end-to-end fate-sharing
by ensuring that the new Semantic Layer retains all end-to-end ``hard state''
and can restart failed TCP flows.

\section{Related Work}
\label{sec-related}



Prior work has explored general protocol decomposition concepts,
such as cross-layer protocol stack optimization~\cite{
	clark90architectural},
modular composition~\cite{
	hutchinson91xkernel,morris99click},
and protocol compilation~\cite{castelluccia96generating}
We focus in contrast on leveraging protocol decomposition
to address the specific problem
of supporting in-path flow optimizers cleanly.

Flow splitting is closely related to TCP splitting~\cite{
	yavatkar94improving,bakre97implementation,rfc3135},
retaining the simplicity, generality, and modularity of TCP splitting
without interfering with end-to-end security or semantics.
Many optimization techniques attempt to avoid
breaking TCP's end-to-end semantics by silently manipulating
a congestion control loop ``from the middle''~\cite{
	balakrishnan95improving,brown97mtcp},
but risk unexpected interactions with other PEPs on the path
or with upgraded endpoints~\cite{vangala03tcp},
and remain incompatible with end-to-end IPsec~\cite{rfc3135},
as described in Section~\ref{sec-patching}.


Like \tng's Flow Layer,
prior work has factored congestion control for other reasons:
TCP control block interdependence~\cite{rfc2140},
Connection Manager~\cite{balakrishnan99integrated},
and TCP/SPAND~\cite{zhang00speeding}
aggregate congestion state across flows,
and DCCP~\abcite{rfc4340}{kohler06dccp} provides
an unreliable, congestion-controlled datagram transport.
DCCP and CM have features
that complement our Flow Layer,
such as CM's support for state aggregation
and application-layer framing~\cite{clark90architectural},
and DCCP's congestion control scheme negotiation.
Other experimental transports
such as 
	Split-TCP~\cite{kopparty02split},	
	pTCP~\cite{hsieh02ptcp},
	mTCP~\cite{zhang04transport},
	LS-SCTP~\cite{al04ls-sctp},		
	and SST~\cite{ford07structured}
have factored congestion control from transport semantics internally
for other reasons.

\xxx{	Moors, "A critical review of “End-to-end arguments in system design”"
	Heimlicher, "The Transport Layer Revisited"}


\com{
Previous work has addressed evolvability challenges
in the Network Layer
through virtualization~\cite{peterson04overcoming}
and anycast routing~\cite{ratnasamy05towards},
but our focus is on the Transport Layer.
}

\tng's Endpoint Layer,
which factors and exposes application endpoint identities to the network,
has precedent in 
Xerox Pup~\cite{boggs80pup}
and AppleTalk~\cite{sidhu90inside},
which include ``socket numbers'' in their network-layer addresses,
and Sirpent~\cite{cheriton89sirpent},
which treats application-level endpoints
as part of Network Layer source routes.
While IP's splitting of endpoint identity across layers
is consistent with the OSI model~\cite{zimmermann80osi},
Tennenhouse argued against layered multiplexing
due to the difficulty it presents
to real-time scheduling~\cite{tennenhouse89layered},
and Feldmeier elaborated on related
issues~\cite{feldmeier90multiplexing}.
Much prior work has focused on firewalls and NATs,
such as NAT traversal schemes~\cite{
	ford05p2p,guha05characterization,biggadike05natblaster},
signaling protocols~\abcite{
	upnp01igd,cheshire05nat}{rfc5190,stiemerling08nat},
and NAT-friendly routing architectures~\cite{
	walfish04middleboxes,guha04nutss}.
We expect that future work exploring \tng's Endpoint Layer 
will draw heavily from this body of work.

\com{
An analogous design with CIDR addressing\abbr{}{~\cite{rfc1518}}
would be to assign each physical host or network interface
a whole ``virtual subnet'' of addresses
identifying individual endpoints on that physical host.
}


\Tng's Isolation Layer is inspired by
location-independent addressing systems
such as 
SFS~\cite{mazieres99separating},
$i3$~\cite{stoica02internet},
HIP~\cite{rfc4423}, and
and UIA~\cite{ford06persistent},
and by IPsec's application-transparent security~\cite{rfc4301};
\tng's contribution is to position such mechanisms
so as to avoid interference
with either the network-oriented or application-oriented functions
of traditional transports.


\section{Conclusion}
\label{sec-concl}

Driven by the challenges of optimizing Internet performance
over today's explosive diversity of network technologies,
the booming network acceleration industry
grew in the US from \$236 million in 2005~\cite{hall06wan}
to \$1 billion in 2009~\cite{mcgillicuddy09wan},
and now markets PEPs implementing a variety of transport-
and higher-level acceleration techniques.
\com{
While this industry markets a variety of acceleration technologies
that operate at multiple levels,
the TCP acceleration functions of devices
sold by companies such as Cisco~\cite{cisco04rate},
	Riverbed~\cite{riverbed08rios},
	and Silver Peak~\cite{silverpeak09overview}
are based on TCP splitting and/or mid-loop tuning.
}
If conventional transport layer PEPs proliferate
like firewalls and NATs already have,
we predict that:
(a)	new transports and end-to-end IPsec
	will become practically undeployable
	{\em even with UDP encapsulation} for NAT/firewall traversal,
	because they will perform poorly on heterogeneous paths
	that optimize only TCP and not UDP traffic; and
(b)	multiple independent mid-loop tuning PEPs will increasingly 
	be found accidentally cohabiting the same TCP paths,
	causing unpredictable control interactions
	and mysterious network failures.

By factoring congestion control to support flow splitting,
\tng demonstrates an architecturally clean alternative to conventional PEPs,
providing the simplicity and generality of TCP splitting,
but without risking unpredictable interactions among mid-loop tuning PEPs,
and without interfering with end-to-end transport-neutral security,
end-to-end semantics, or fate-sharing.
While we make no pretense that this paper 
defines a complete
next-generation transport services architecture,
or that flow splitting alone
would drive the widespread deployment of such an architecture,
we hope that the many benefits potentially achievable at once
from a careful factoring of congestion control from transport semantics~\cite{
	rfc2140,balakrishnan99integrated,zhang00speeding,
	al04ls-sctp,ford07structured}
will eventually drive the deployment
of a next-generation architecture incorporating these ideas.

\com{
\subsection*{Acknowledgments}

XXX Stuart Cheshire, ...
}

\begin{scriptsize}
\bibliography{os}

\begin{thebibliography}{100}

\bibitem{aiello04jfk}
W.~Aiello et~al.
\newblock Just fast keying: Key agreement in a hostile {Internet}.
\newblock {\em \bibbrev{TISSEC}{ACM Transactions on Information and System
  Security (TISSEC)}}, 7(2):1--32, May 2004.

\bibitem{akyildiz01tcp}
I.~F. Akyildiz, G.~Morabito, and S.~Palazzo.
\newblock {TCP-Peach}: A new congestion control scheme for satellite {IP}
  networks.
\newblock {\em Transactions on Networking}, 9(3), June 2001.

\bibitem{al04ls-sctp}
A.~A.~E. Al, T.~Saadawi, and M.~Lee.
\newblock {LS-SCTP}: a bandwidth aggregation technique for stream control
  transmission protocol.
\newblock {\em Computer Communications}, 27(10):1012--1024, June 2004.

\bibitem{allman96application}
M.~Allman, H.~Kruse, and S.~Ostermann.
\newblock An application-level solution to {TCP}'s satellite inefficiencies.
\newblock In {\em 1st \bibbrev{WOSBIS}{Workshop on Satellite-based Information
  Services (WOSBIS)}}, Nov. 1996.

\bibitem{rfc2581}
M.~Allman, V.~Paxson, and W.~Stevens.
\newblock {TCP} congestion control, Apr. 1999.
\newblock RFC 2581.

\bibitem{baiocchi07yeah}
A.~Baiocchi, A.~P. Castellani, and F.~Vacirca.
\newblock {YeAH-TCP}: Yet another highspeed {TCP}.
\newblock In {\em 5th \bibbrev{PFLDnet Workshop}{Workshop on Protocols for Fast
  Long-Distance Networks (PFLDnet)}}, Feb. 2007.

\bibitem{rfc1812}
F.~{Baker, ed.}
\newblock Requirements for {IP} version 4 routers, June 1995.
\newblock RFC 1812.

\bibitem{bakre97implementation}
A.~V. Bakre and B.~Badrinath.
\newblock Implementation and performance evaluation of indirect {TCP}.
\newblock {\em IEEE Transactions on Computers}, 46(3):260--278, Mar. 1997.

\bibitem{balakrishnan98explicit}
H.~Balakrishnan and R.~H. Katz.
\newblock Explicit loss notification and wireless web performance.
\newblock In {\em IEEE Globecom Internet Mini-Conference}, Nov. 1998.

\bibitem{balakrishnan99integrated}
H.~Balakrishnan, H.~S. Rahul, and S.~Seshan.
\newblock An integrated congestion management architecture for {Internet}
  hosts.
\newblock In {\em \bibbrev{SIGCOMM}{ACM SIGCOMM}}, Sept. 1999.

\bibitem{balakrishnan95improving}
H.~Balakrishnan, S.~Seshan, E.~Amir, and R.~H. Katz.
\newblock Improving {TCP/IP} performance over wireless networks.
\newblock In {\em 1st \bibbrev{MOBICOM}{International Conference on Mobile
  Computing and Networking (MOBICOM)}}, Nov. 1995.

\bibitem{barakat99tcp}
C.~Barakat, E.~Altman, and W.~Dabbous.
\newblock On {TCP} performance in an heterogeneous network: A survey.
\newblock Technical Report 3737, INRIA, July 1999.

\bibitem{rfc1948}
S.~Bellovin.
\newblock Defending against sequence number attacks, May 1996.
\newblock RFC 1948.

\bibitem{biggadike05natblaster}
A.~Biggadike et~al.
\newblock {NATBLASTER}: Establishing {TCP} connections between hosts behind
  {NATs}.
\newblock In {\em ACM SIGCOMM Asia Workshop}, Apr. 2005.

\bibitem{boggs80pup}
D.~R. Boggs, J.~F. Shoch, E.~A. Taft, and R.~M. Metcalfe.
\newblock Pup: An internetwork architecture.
\newblock {\em IEEE Transactions on Communications}, 28(4):612--624, Apr. 1980.

\bibitem{rfc3135}
J.~Border et~al.
\newblock Performance enhancing proxies intended to mitigate link-related
  degradations, June 2001.
\newblock RFC 3135.

\bibitem{rfc1122}
R.~{Braden, ed.}
\newblock Requirements for {Internet} hosts --- communication layers, Oct.
  1989.
\newblock RFC 1122.

\bibitem{brakmo95tcp}
L.~Brakmo and L.~Peterson.
\newblock {TCP Vegas}: End to end congestion avoidance on a global {Internet}.
\newblock {\em IEEE Journal on Selected Areas in Communications},
  13(8):1465--1480, Oct. 1995.

\bibitem{brown97mtcp}
K.~Brown and S.~Singh.
\newblock {M-TCP}: {TCP} for mobile cellular networks.
\newblock {\em Computer Communications Review}, 27(5):19--43, Oct. 1997.

\bibitem{caceres95improving}
R.~C\'aceres and L.~Iftode.
\newblock Improving the performance of reliable transport protocols in mobile
  computing environments.
\newblock {\em IEEE Journal on Selected Areas in Communications},
  13(5):850--857, June 1995.

\bibitem{casetti02tcp}
C.~Casetti, M.~Gerla, S.~Mascolo, M.~Sanadidi, and R.~Wang.
\newblock {TCP Westwood}: End-to-end congestion control for wired/wireless
  networks.
\newblock {\em Wireless Networks}, 8(5):467--479, Sept. 2002.

\bibitem{castelluccia96generating}
C.~Castelluccia and W.~Dabbous.
\newblock Generating efficient protocol code from an abstract specification.
\newblock In {\em \bibbrev{SIGCOMM}{ACM SIGCOMM}}, Aug. 1996.

\bibitem{cheriton89sirpent}
D.~R. Cheriton.
\newblock Sirpent: A high-performance internetworking approach.
\newblock In {\em \bibbrev{SIGCOMM}{ACM SIGCOMM}}, Sept. 1989.

\bibitem{cheshire05nat}
S.~Cheshire, M.~Krochmal, and K.~Sekar.
\newblock {NAT} port mapping protocol, June 2005.
\newblock Internet-Draft (Work in Progress).

\bibitem{chockalingam99wireless}
A.~Chockalingam, M.~Zorzi, and V.~Tralli.
\newblock Wireless {TCP} performance with link layer {FEC/ARQ}.
\newblock In {\em \bibconf{ICC}{IEEE International Conference on
  Communications}}, June 1999.

\bibitem{cisco04rate}
{Cisco, Inc.}
\newblock Rate based satellite control protocol, 2004.

\bibitem{clark88design}
D.~D. Clark.
\newblock The design philosophy of the {DARPA} {Internet} protocols.
\newblock In {\em \bibbrev{SIGCOMM}{ACM SIGCOMM}}, Aug. 1988.

\bibitem{clark90architectural}
D.~D. Clark and D.~L. Tennenhouse.
\newblock Architectural considerations for a new generation of protocols.
\newblock In {\em \bibbrev{SIGCOMM}{ACM SIGCOMM}}, pages 200--208, 1990.

\bibitem{crowcroft98differentiated}
J.~Crowcroft and P.~Oechslin.
\newblock Differentiated end-to-end internet services using a weighted
  proportional fair sharing {TCP}.
\newblock {\em ACM \bibbrev{CCR}{Computer Communications Review}},
  28(3):53--69, July 1998.

\bibitem{davies72control}
D.~W. Davies.
\newblock The control of congestion in packet switching networks.
\newblock {\em IEEE Transactions on Communications}, 20(3):546--550, June 1972.

\bibitem{rfc4346}
T.~Dierks and E.~Rescorla.
\newblock The transport layer security {(TLS)} protocol version 1.1, Apr. 2006.
\newblock RFC 4346.

\bibitem{dischinger07characterizing}
M.~Dischinger, A.~Haeberlen, K.~P. Gummadi, and S.~Saroiu.
\newblock Characterizing residential broadband networks.
\newblock In {\em \bibbrev{IMC}{Internet Measurement Conference (IMC)}}, Oct.
  2007.

\bibitem{rfc4987}
W.~Eddy.
\newblock {TCP} {SYN} flooding attacks and common mitigations, Aug. 2007.
\newblock RFC 4987.

\bibitem{feldmeier90multiplexing}
D.~C. Feldmeier.
\newblock Multiplexing issues in communication system design.
\newblock In {\em \bibbrev{SIGCOMM}{ACM SIGCOMM}}, Sept. 1990.

\bibitem{ferrill06network}
P.~Ferrill.
\newblock Network traffic shaping tools.
\newblock {\em Processor}, 28(16):4, Apr. 2006.

\bibitem{finn89connectionless}
G.~G. Finn.
\newblock A connectionless congestion control algorithm.
\newblock {\em ACM \bibbrev{CCR}{Computer Communications Review}},
  19(5):12--31, Oct. 1989.

\bibitem{floyd91connections}
S.~Floyd.
\newblock Connections with multiple congested gateways in packet-switched
  networks, part 1: One-way traffic.
\newblock {\em ACM \bibbrev{CCR}{Computer Communications Review}},
  21(5):30--47, Oct. 1991.

\bibitem{rfc3649}
S.~Floyd.
\newblock {HighSpeed TCP} for large congestion windows, Dec. 2003.
\newblock RFC 3649.

\bibitem{floyd99promoting}
S.~Floyd and K.~Fall.
\newblock Promoting the use of end-to-end congestion control in {the Internet}.
\newblock {\em Transactions on Networking}, 7(4):458--472, Aug. 1999.

\bibitem{floyd93random}
S.~Floyd and V.~Jacobson.
\newblock Random early detection gateways for congestion avoidance.
\newblock {\em Transactions on Networking}, 1(4):1063--6692, Aug. 1993.

\bibitem{ford05p2p}
B.~Ford.
\newblock Peer-to-peer communication across network address translators.
\newblock In {\em \bibbrev{USENIX}{USENIX Annual Technical Conference}}, Apr.
  2005.

\bibitem{ford07structured}
B.~Ford.
\newblock Structured streams: a new transport abstraction.
\newblock In {\em \bibbrev{SIGCOMM}{ACM SIGCOMM}}, Aug. 2007.

\bibitem{ford06persistent}
B.~Ford et~al.
\newblock Persistent personal names for globally connected mobile devices.
\newblock In {\em 7th \bibbrev{OSDI}{USENIX Symposium on Operating Systems
  Design and Implementation (OSDI)}}, Nov. 2006.

\bibitem{ford08breaking}
B.~Ford and J.~Iyengar.
\newblock Breaking up the transport logjam.
\newblock In {\em \bibbrev{HotNets-VII}{7th Workshop on Hot Topics in Networks
  (HotNets-VII)}}, Oct. 2008.

\bibitem{rfc2979}
N.~Freed.
\newblock Behavior of and requirements for {Internet} firewalls, Oct. 2000.
\newblock RFC 2979.

\bibitem{gerla80flow}
M.~Gerla and L.~Kleinrock.
\newblock Flow control : A comparative survey.
\newblock {\em IEEE Transactions on Communications}, 28(4):553--574, Apr. 1980.

\bibitem{guha05characterization}
S.~Guha and P.~Francis.
\newblock Characterization and measurement of {TCP} traversal through {NATs}
  and firewalls.
\newblock In {\em \bibbrev{IMC}{Internet Measurement Conference (IMC)}}, Oct.
  2005.

\bibitem{guha04nutss}
S.~Guha, Y.~Takeday, and P.~Francis.
\newblock {NUTSS}: A {SIP}-based approach to {UDP} and {TCP} network
  connectivity.
\newblock In {\em SIGCOMM 2004 Workshops}, Aug. 2004.

\bibitem{habib01unresponsive}
A.~Habib and B.~Bhargava.
\newblock Unresponsive flow detection and control using the differentiated
  services framework.
\newblock In {\em \bibbrev{PDCS}{IASTED International Conference on Parallel
  and Distributed Computing Systems (PDCS)}}, Aug. 2001.

\bibitem{hall06wan}
M.~Hall.
\newblock {WAN} optimization dominated by startups, growing fast.
\newblock {\em Enterprise Networking Planet}, Apr. 2006.

\bibitem{holland02analysis}
G.~Holland and N.~Vaidya.
\newblock Analysis of {TCP} performance over mobile ad hoc networks.
\newblock {\em Wireless Networks}, 8(2), Mar. 2002.

\bibitem{rfc3686}
R.~Housley.
\newblock Using advanced encryption standard {(AES)} counter mode with {IPsec}
  encapsulating security payload {(ESP)}, Jan. 2004.
\newblock RFC 3686.

\bibitem{hsieh02ptcp}
H.-Y. Hsieh and R.~Sivakumar.
\newblock {pTCP}: An end-to-end transport layer protocol for striped
  connections.
\newblock In {\em \bibconf[10th]{ICNP}{International Conference on Network
  Protocols}}, Nov. 2002.

\bibitem{hutchinson91xkernel}
N.~C. Hutchinson and L.~L. Peterson.
\newblock The {x-Kernel}: An architecture for implementing network protocols.
\newblock {\em IEEE Transactions on Software Engineering}, 17(1), Jan. 1991.

\bibitem{inamura04impact}
H.~Inamura et~al.
\newblock Impact of layer two {ARQ} on {TCP} performance in {W-CDMA} networks.
\newblock In {\em \bibbrev{ICDCS}{International Conference on Distributed
  Computing Systems}}, Mar. 2004.

\bibitem{jacobson88congestion}
V.~Jacobson.
\newblock Congestion avoidance and control.
\newblock pages 314--329, Aug. 1988.

\bibitem{jin99spack}
K.~Jin, K.~Kim, and J.~Lee.
\newblock {SPACK}: rapid recovery of the {TCP} performance using split-ack in
  mobile communication environments.
\newblock In {\em IEEE Region 10 Conference}, Sept. 1999.

\bibitem{jin03spectrum}
S.~Jin et~al.
\newblock A spectrum of {TCP}-friendly window-based congestion control
  algorithms.
\newblock {\em Transactions on Networking}, 11(3):341--355, June 2003.

\bibitem{katabi02internet}
D.~Katabi, M.~Handley, and C.~Rohrs.
\newblock Internet congestion control for high bandwidth-delay product
  networks.
\newblock In {\em \bibbrev{SIGCOMM}{ACM SIGCOMM}}, Aug. 2002.

\bibitem{rfc4306}
C.~{Kaufman, Ed.}
\newblock Internet key exchange {(IKEv2)} protocol, Dec. 2005.
\newblock RFC 4306.

\bibitem{rfc4868}
S.~Kelly and S.~Frankel.
\newblock Using {HMAC-SHA-256}, {HMAC-SHA-384}, and {HMAC-SHA-512} with
  {IPsec}, May 2007.
\newblock RFC 4868.

\bibitem{kelly03scalable}
T.~Kelly.
\newblock Scalable {TCP}: Improving performance in highspeed wide area
  networks.
\newblock {\em Computer Communications Review}, 33(2):83--91, Apr. 2003.

\bibitem{rfc4301}
S.~Kent and K.~Seo.
\newblock Security architecture for the {Internet} protocol, Dec. 2005.
\newblock RFC 4301.

\bibitem{rfc4340}
E.~Kohler, M.~Handley, and S.~Floyd.
\newblock Datagram congestion control protocol {(DCCP)}, Mar. 2006.
\newblock RFC 4340.

\bibitem{kopparty02split}
S.~Kopparty, S.~V. Krishnamurthy, M.~Faloutsos, and S.~K. Tripathi.
\newblock Split {TCP} for mobile ad hoc networks.
\newblock In {\em IEEE GLOBECOM}, Nov. 2002.

\bibitem{kung93fcvc}
H.~T. Kung and A.~Chapman.
\newblock The {FCVC} (flow-controlled virtual channels) proposal for {ATM}
  networks: A summary.
\newblock In {\em \bibconf[1st]{ICNP}{International Conference on Network
  Protocols}}, Oct. 1993.

\bibitem{liu01atcp}
J.~Liu and S.~Singh.
\newblock {ATCP}: {TCP} for mobile ad hoc networks.
\newblock {\em IEEE Journal on Selected Areas in Communications},
  19(7):1300--1315, July 2001.

\bibitem{lochert07survey}
C.~Lochert, B.~Scheuermann, and M.~Mauve.
\newblock A survey on congestion control for mobile ad-hoc networks.
\newblock {\em \bibbrev{WCMC}{Wireless Communications \& Mobile Computing}},
  7(5):655--676, June 2007.

\bibitem{magalhaes01transport}
L.~Magalhaes and R.~Kravets.
\newblock Transport level mechanisms for bandwidth aggregation on mobile hosts.
\newblock In {\em \bibconf[9th]{ICNP}{International Conference on Network
  Protocols}}, Nov. 2001.

\bibitem{mazieres99separating}
D.~Mazi{\`e}res, M.~Kaminsky, M.~F. Kaashoek, and E.~Witchel.
\newblock {S}eparating key management from file system security.
\newblock In {\em 17th \bibbrev{SOSP}{Symposium on Operating System
  Principles}}, Dec. 1999.

\bibitem{mcgillicuddy09wan}
S.~McGillicuddy.
\newblock {WAN} optimization market passes \$1 billion; {Cisco} takes the lead.
\newblock {\em SearchEnterpriseWAN.com}, Mar. 2009.

\bibitem{mishra92hopbyhop}
P.~P. Mishra and H.~Kanakia.
\newblock A hop by hop rate-based congestion control scheme.
\newblock In {\em \bibbrev{SIGCOMM}{ACM SIGCOMM}}, Aug. 1992.

\bibitem{mo99analysis}
J.~Mo, R.~J. La, V.~Anantharam, and J.~Walrand.
\newblock Analysis and comparison of {TCP Reno} and {Vegas}.
\newblock In {\em \bibbrev{INFOCOM}{IEEE INFOCOM}}, Mar. 1999.

\bibitem{morris99click}
R.~Morris, E.~Kohler, J.~Jannotti, and M.~F. Kaashoek.
\newblock The {Click} modular router.
\newblock In {\em 17th \bibbrev{SOSP}{ACM Symposium on Operating System
  Principles}}, Dec. 1999.

\bibitem{rfc5201}
R.~Moskowitz et~al.
\newblock Host identity protocol, Apr. 2008.
\newblock RFC 5201.

\bibitem{rfc4423}
R.~Moskowitz and P.~Nikander.
\newblock Host identity protocol {(HIP)} architecture, May 2006.
\newblock RFC 4423.

\bibitem{rfc896}
J.~Nagle.
\newblock {Congestion Control in IP/TCP Internetworks}, Jan. 1984.
\newblock RFC 896.

\bibitem{rfc1151}
C.~Partridge and R.~Hinden.
\newblock Version 2 of the reliable data protocol {(RDP)}, Apr. 1990.
\newblock RFC 1151.

\bibitem{rfc768}
J.~Postel.
\newblock User datagram protocol, Aug. 1980.
\newblock RFC 768.

\bibitem{rfc1016}
W.~Prue and J.~Postel.
\newblock Something a host could do with source quench: The source quench
  introduced delay {(SQuID)}, July 1987.
\newblock RFC 1016.

\bibitem{rfc3168}
K.~Ramakrishnan, S.~Floyd, and D.~Black.
\newblock The addition of explicit congestion notification {(ECN)} to {IP},
  Sept. 2001.
\newblock RFC 3168.

\bibitem{rangarajan99eruf}
A.~Rangarajan and A.~Acharya.
\newblock {ERUF}: Early regulation of unresponsive best-effort traffic.
\newblock In {\em \bibconf[7th]{ICNP}{International Conference on Network
  Protocols}}, Oct. 1999.

\bibitem{rfc4347}
E.~Rescorla and N.~Modadugu.
\newblock Datagram transport layer security, Apr. 2006.
\newblock RFC 4347.

\bibitem{roberts03next}
L.~G. Roberts.
\newblock The next generation of {IP} --- flow routing.
\newblock In {\em \bibbrev{SSGRR}{International Conference on Advances in
  Infrastructure for Electronic Business, Science, Education, Medicine, and
  Mobile Technologies on the Internet}}, July 2003.

\bibitem{rosenberg08udp}
J.~Rosenberg.
\newblock {UDP} and {TCP} as the new waist of the {Internet} hourglass, Feb.
  2008.
\newblock Internet-Draft (Work in Progress).

\bibitem{saltzer84endtoend}
J.~H. Saltzer, D.~P. Reed, and D.~D. Clark.
\newblock End-to-end arguments in system design.
\newblock {\em \bibbrev{TOCS}{Transactions on Computer Systems}},
  2(4):277--288, Nov. 1984.

\bibitem{savage99tcp}
S.~Savage et~al.
\newblock {TCP} congestion control with a misbehaving receiver.
\newblock {\em Computer Communications Review}, 29(5), Oct. 1999.

\bibitem{sidhu90inside}
G.~S. Sidhu, R.~F. Andrews, and A.~B. Oppenheimer.
\newblock {\em Inside {Appletalk}}.
\newblock Addison-Wesley, 2rd edition, 1990.

\bibitem{sinha02wtcp}
P.~Sinha et~al.
\newblock {WTCP}: A reliable transport protocol for wireless wide-area
  networks.
\newblock {\em Wireless Networks}, 8(2):301--316, Mar. 2002.

\bibitem{sivakumar00psockets}
H.~Sivakumar, S.~Bailey, and R.~Grossman.
\newblock {PSockets}: The case for application-level network striping for data
  intensive applications using high speed wide area networks.
\newblock In {\em \bibbrev{SC2000}{Conference on High Performance Networking
  and Computing (SC2000)}}, Nov. 2000.

\bibitem{rfc3022}
P.~Srisuresh and K.~Egevang.
\newblock Traditional {IP} network address translator {(Traditional NAT)}, Jan.
  2001.
\newblock RFC 3022.

\bibitem{stanojevic08can}
M.~Stanojevic, R.~Mahajan, T.~Millstein, and M.~Musuvathi.
\newblock Can you fool me? towards automatically checking protocol gullibility.
\newblock In {\em \bibbrev{HotNets-VII}{7th Workshop on Hot Topics in Networks
  (HotNets-VII)}}, Oct. 2008.

\bibitem{rfc4960}
R.~{Stewart, ed.}
\newblock Stream control transmission protocol, Sept. 2007.
\newblock RFC 4960.

\bibitem{stoica02internet}
I.~Stoica et~al.
\newblock Internet indirection infrastructure.
\newblock In {\em \bibbrev{SIGCOMM}{ACM SIGCOMM}}, Aug. 2002.

\bibitem{stoica98core}
I.~Stoica, S.~Shenker, and H.~Zhang.
\newblock Core-stateless fair queueing: A scalable architecture to approximate
  fair bandwidth allocations in high speed networks.
\newblock In {\em \bibbrev{SIGCOMM}{ACM SIGCOMM}}, Aug. 1998.

\bibitem{sundaresan03atp}
K.~Sundaresan, V.~Anantharaman, H.~Hsieh, and R.~Sivakumar.
\newblock {ATP}: A reliable transport protocol for ad-hoc networks.
\newblock In {\em \bibbrev{ACM MOBIHOC}{ACM International Symposium on Mobile
  Ad Hoc Networking and Computing (MOBIHOC)}}, June 2003.

\bibitem{swany04improving}
M.~Swany.
\newblock Improving throughput for grid applications with network logistics.
\newblock In {\em \bibbrev{SC2004}{High Performance Computing, Networking and
  Storage Conference (SC2004)}}, Nov. 2004.

\bibitem{tan06compound}
K.~Tan, J.~Song, Q.~Zhang, and M.~Sridharan.
\newblock {Compound TCP}: A scalable and {TCP}-friendly congestion control for
  high-speed networks.
\newblock In {\em \bibbrev{INFOCOM}{IEEE INFOCOM}}, Apr. 2006.

\bibitem{rfc793}
Transmission control protocol, Sept. 1981.
\newblock RFC 793.

\bibitem{tennenhouse89layered}
D.~L. Tennenhouse.
\newblock Layered multiplexing considered harmful.
\newblock In {\em 1st International Workshop on Protocols for High-Speed
  Networks}, May 1989.

\bibitem{rfc2140}
J.~Touch.
\newblock {TCP} control block interdependence, Apr. 1997.
\newblock RFC 2140.

\bibitem{touch06tcp}
J.~Touch.
\newblock A {TCP} option for port names, Apr. 2006.
\newblock Internet-Draft (Work in Progress).

\bibitem{touch06recursive}
J.~D. Touch, Y.-S. Wang, and V.~Pingali.
\newblock A recursive network architecture.
\newblock Technical Report ISI-TR-2006-626, University of Southern California
  Information Sciences Institute, Oct. 2006.

\bibitem{trolltech-qt}
Trolltech.
\newblock Qt cross-platform application framework.
\newblock \url{http://trolltech.com/products/qt/}.

\bibitem{upnp01igd}
{UPnP Forum}.
\newblock Internet gateway device {(IGD)} standardized device control protocol,
  Nov. 2001.
\newblock \url{http://www.upnp.org/}.

\bibitem{vangala03tcp}
S.~Vangala and M.~A. Labrador.
\newblock The {TCP SACK}-aware snoop protocol for {TCP} over wireless networks.
\newblock In {\em Vehicular Technology Conference}, Oct. 2003.

\bibitem{walfish04middleboxes}
M.~Walfish, J.~Stribling, M.~Krohn, H.~Balakrishnan, R.~Morris, and S.~Shenker.
\newblock Middleboxes no longer considered harmful.
\newblock In {\em USENIX Symposium on Operating Systems Design and
  Implementation}, Dec. 2004.

\bibitem{wong99improving}
J.~W. Wong and V.~C. Leung.
\newblock Improving end-to-end performance of {TCP} using link-layer
  retransmissions over mobile internetworks.
\newblock In {\em \bibconf{ICC}{IEEE International Conference on
  Communications}}, June 1999.

\bibitem{yavatkar94improving}
R.~Yavatkar and N.~Bhagawat.
\newblock Improving end-to-end performance of {TCP} over mobile internetworks.
\newblock In {\em Workshop on Mobile Computing Systems and Applications}, Dec.
  1994.

\bibitem{yi07hopbyhop}
Y.~Yi and S.~Shakkottai.
\newblock Hop-by-hop congestion control over a wireless multi-hop network.
\newblock {\em IEEE Transactions on Networking}, 15(1):133--144, Feb. 2007.

\bibitem{zhang04transport}
M.~Zhang, J.~Lai, A.~Krishnamurthy, L.~Peterson, and R.~Wang.
\newblock A transport layer approach for improving end-to-end performance and
  robustness using redundant paths.
\newblock In {\em \bibbrev{USENIX}{USENIX Annual Technical Conference}}, June
  2004.

\bibitem{zhang00speeding}
Y.~Zhang, L.~Qiu, and S.~Keshav.
\newblock Speeding up short data transfers: Theory, architectural support and
  simulation results.
\newblock In {\em \bibconf[10th]{NOSSDAV}{Network and Operating System Support
  for Digital Audio and Video}}, June 2000.

\bibitem{zimmermann80osi}
H.~Zimmermann.
\newblock {OSI} reference model---the {ISO} model of architecture for open
  systems interconnection.
\newblock {\em IEEE Transactions on Communications}, 28(4):425--432, Apr. 1980.

\end{thebibliography}
\bibliographystyle{abbr}
\end{scriptsize}

\end{document}